\documentclass[prd,onecolumn,notitlepage,showpacs,preprintnumbers,amsmath,amssymb,nofootinbib,APS,10pt,superscriptaddress]{revtex4-2}

\usepackage{dcolumn}
\usepackage{bm}
\usepackage[dvipsnames]{xcolor}
\usepackage[spanish,english]{babel}
\usepackage[utf8]{inputenc}
\usepackage{amsmath,amssymb,amsfonts,latexsym,cancel}
\usepackage{graphicx}
\usepackage{svg}
\usepackage{color}
\usepackage{soul}
\usepackage[normalem]{ulem}
\usepackage{hyperref}
\usepackage{amsmath}
\usepackage{slashed}
\usepackage{comment}
\usepackage{subfig}
\usepackage{float} 
\usepackage{mathtools}
\usepackage{mathrsfs}
\usepackage{braket}
\usepackage{enumitem}
\usepackage[many]{tcolorbox}

\allowdisplaybreaks

\newcommand{\be}{\begin{equation}}
\newcommand{\ee}{\end{equation}}
\newcommand{\bea}{\begin{eqnarray}}
\newcommand{\eea}{\end{eqnarray}}
\newcommand{\mH}{\mathcal{H}}
\newcommand{\hr}{\hat{\rho}}
\newcommand{\ha}{\hat{a}}
\newcommand{\had}{\hat{a}^\dagger}
\newcommand{\hx}{\hat{x}}
\newcommand{\hp}{\hat{p}}
\newcommand{\hR}{\hat{R}}
\newcommand{\bs}{\bm{\sigma}}
\newcommand{\Tr}{\text{Tr}}

\newcommand{\pr}{\prime}
\newcommand{\tin}{\text{in}}
\newcommand{\tout}{\text{out}}

\begin{document}

\preprint{ TUM-HEP-1582/25}
\title{{\bf Entanglement in the Schwinger effect}}

\author{Dimitrios Kranas}\email{dimitrioskranas@gmail.com}

\affiliation{Laboratoire de Physique de l’Ecole Normale Supérieure,
ENS, CNRS, Université PSL, Sorbonne Université,
Université Paris Cité, 75005 Paris, France}

\affiliation{Universidad Carlos III de Madrid, Departamento de Matem\'{a}ticas.
Avenida de la Universidad 30 (edificio Sabatini), 28911 Legan\'{e}s (Madrid), Spain}

\author{Amaury Marchon}\email{amaury.marchon@lapth.cnrs.fr}

\affiliation{ENS de Lyon, CNRS, Laboratoire de Physique, F-69342 Lyon, France}

\affiliation{Physik-Department, Technische Universität München, James-Franck-Str., 85748 Garching, Germany}

\affiliation{Laboratoire d'Annecy de Physique Théorique, CNRS - USMB, BP 110 Annecy-le-Vieux, 74941 Annecy, France}

\author{Silvia Pla}\email{silvia.pla-garcia@tum.de}

\affiliation{Physik-Department, Technische Universität München, James-Franck-Str., 85748 Garching, Germany}

\date{\today}

\begin{abstract}
We analyze entanglement generated by the Schwinger effect using a mode-by-mode formalism for scalar and spinor QED in constant backgrounds. Starting from thermal initial states, we derive compact, closed-form results for bipartite entanglement between particle–antiparticle partners in terms of the Bogoliubov coefficients. For bosons, thermal fluctuations enhance production but suppress quantum correlations: the logarithmic negativity is nonzero only below a (mode-dependent) critical temperature $T_c$. At fixed $T$, entanglement appears only above a critical field $E_{\text{entang}}$. For fermions, we observe a qualitatively different pattern:  the fermionic logarithmic negativity is non-vanishing at finite temperature, and is monotonically suppressed by thermal noise. As a function of the electric field, it is non-monotonic, featuring a temperature-independent optimal field strength $E_*$ and decreasing on both sides of the maximum. We give quantitative estimates for analog experiments, where our entanglement criteria convert directly into concrete temperature and electric field constraints. These findings identify realistic regimes where the quantum character of Schwinger physics may be tested in the laboratory. 

\end{abstract}

\maketitle

\tableofcontents

\newpage

\section{Introduction}
\label{sec:intro}

The Schwinger effect -- the spontaneous creation of particle-antiparticle pairs from the vacuum in the presence of a strong electric field -- is one of the most distinctive predictions of quantum field theory under extreme conditions. Originally formulated in the context of quantum electrodynamics (QED) \cite{Schwinger:1951nm}, it shares deep connections with other interesting phenomena such as cosmological particle production \cite{Parker:2025jef,Parker:1968mv,Parker:1969au} and the Hawking effect \cite{Hawking:1974rv,Hawking:1975vcx}. These processes not only signal an instability of the vacuum, but also imprint non-trivial quantum correlations in the created particles, which can be used, in principle, to probe the quantum nature of the underlying dynamics.\\

On the experimental side, there is growing interest in experimentally probing non-perturbative effects in QED. High-intensity laser facilities aim to reach the critical field strength where vacuum pair production becomes observable \cite{Dunne:2008kc,Karbstein:2019oej,Fedotov:2022ely}. At the same time, analogue experiments have emerged as powerful tools to simulate aspects of strong-field QED in condensed matter systems. The mesoscopic Schwinger effect provides an excellent playground to study properties of the (analogue) Schwinger effect for fermions, since a universal Schwinger conductance in graphene was reported \cite{Schmitt:2022pkd}. For bosons, magnetic inhomogeneities in chiral magnets have been proposed as a route to Schwinger-like magnon production \cite{Hongo:2020xaw,Adorno:2023olb}.\footnote{For complementary perspectives on entanglement in Schwinger pair creation in holographic setups, see \cite{Grieninger:2023ehb,Grieninger:2023pyb}, building on the seminal proposals \cite{Jensen:2013ora,Sonner:2013mba}.} \\ 

The Schwinger effect at finite temperature has also been studied extensively from the viewpoint of the thermal effective action. For purely magnetic backgrounds, the corresponding thermal effective action was studied for the first time in Ref.~\cite{Dittrich:1979ux}. Extensions to general constant electromagnetic backgrounds  were considered, for example, in Refs.~\cite{Loewe:1992teft,Gies:1998vt}. More recently, thermal Schwinger pair production has been revisited within the worldline formalism in Refs.~\cite{Medina:2015qzc,Gould:2018ovk,Korwar:2018euc}, where one-loop thermal corrections, threshold effects, the high-temperature (worldline-sphaleron) regime, and extensions to coexistent electric and magnetic fields have been discussed.\\

In this work, we explore the Schwinger effect through the lens of mode-by-mode entanglement, with the aim of understanding it from a quantum information perspective. We study both scalar and spinor QED in the external field approximation, and make use of  the Gaussian formalism to capture the structure of bipartite entanglement between modes. This approach allows us to describe how particle–antiparticle entanglement is generated during pair production and to characterize the entanglement profile in momentum space. We perform a systematic comparison between the bosonic and fermionic cases, identifying both qualitative and quantitative differences in their entanglement profile. \\

Our analysis is performed in the external-field approximation, with the electromagnetic field treated as a prescribed classical background and the charged sector described by a quadratic Hamiltonian. We consider a real-time evolution driven by the Hamiltonian and encoded in a Bogoliubov transformation, from an initially unentangled state, such as the vacuum or a thermal state, to a final state that generically contains particle-antiparticle entanglement. By construction, this framework does not include thermal modifications of the electromagnetic background, such as Debye screening, nor the backreaction of the produced charges on the external field. Our results should therefore be interpreted as characterizing the charged matter sector in prescribed backgrounds.\\

Building on this framework, we show 
that the particle-antiparticle pairs generated via vacuum polarization necessarily carry entanglement that cannot be reproduced by classical stochastic processes or thermal noise. Moreover, we quantify how much this entanglement degrades  when the initial state is thermal, identifying a {\it critical temperature} above which the  (bosonic) logarithmic negativity vanishes, even though pair production persists. This provides an operational criterion to distinguish the quantum nature of the process. For fermions, we  employ the fermionic logarithmic negativity proposed in Refs.   \cite{Shapourian:2016cqu,Shapourian:2018ozl} and find that entanglement is always present at finite temperature, although it is strongly suppressed as thermal noise increases. \\

At a fixed temperature, we also analyze entanglement in terms of the electric field. For bosons, there is only entanglement if the electric field is above a critical value $E_{\text{entang}}$. For fermions, the pattern is different: the logarithmic negativity is a non-monotonic function of the electric field, featuring a temperature-independent optimal field strength $E_{*}$, and decays for bigger and smaller values of $E$.
We also study a mechanism -- valid for scalar fields -- to enhance entanglement by tuning the initial state. Specifically, we show that by preparing the system in a squeezed state, one can increase the quantum entanglement produced during the pair creation process of bosons. This provides a way to control quantum correlations in strong-field pair production, with potential applications in analogue experiments. \\

The entanglement structure of pairs produced via the Schwinger mechanism has been previously studied in various settings, using different entanglement measures such as the von Neumann entropy, and bosonic logarithmic negativity. Previous works have focused on both scalar and spinor fields in constant or pulsed backgrounds at zero temperature \cite{Ebadi2014,Li_2017,Bhattacharya2020,Kaushal_2025}. Related links between entropy production and entanglement in electric-field quenches and pulses have also been discussed recently \cite{Florio:2021xvj,Dunne:2022zlx}, while the interplay between entanglement and thermalization in the massive Schwinger model has been recently explored in Ref.~\cite{Florio:2025hoc}.  In this work, we investigate how different input states, such as thermal or single-mode squeezed states, affect the entanglement generated in the Schwinger effect, thus providing a  broad analysis of the quantum nature of the Schwinger process. Furthermore, we apply the fermionic logarithmic negativity using the separability criterion established in Ref. \cite{Shapourian:2016cqu}. Beyond strong QED, recent efforts in collider physics \cite{Afik:2025ejh,Aguilar-Saavedra:2022uye,Aguilar-Saavedra:2022wam} and early-universe cosmology \cite{Pueyo:2024twm,Bhardwaj:2023squ} have explored how entanglement-based observables can be used to test quantum correlations in fundamental processes. Similar ideas have been recently discussed in the context of gravitational radiation \cite{Manikandan:2025qgv,Manikandan:2025hlz}.\\

The outline of the article is as follows. In Section \ref{sec:Schwinger} we briefly review the Schwinger effect for both scalars and fermions following a mode-by-mode approach in terms of the Bogoliubov coefficients. This approach is particularly useful for understanding quantum entanglement. We focus on two cases: i) constant electric field $\vec E$,  and ii) constant and anti-parallel electric and magnetic fields $\vec E$ and $\vec B$. We then introduce the tools needed to quantify quantum entanglement, including the Gaussian formalism, and the definition of logarithmic negativity as our main measure of entanglement in Section \ref{sec:entanglement}. In Section \ref{sec:mainresults} we present our main results: a comparative study of scalar and spinor entanglement in QED, the impact of thermal initial states, and the identification of critical temperatures and electric field strengths above (or under) which entanglement vanishes (for bosons) or is strongly suppressed (for fermions). We also discuss how these findings can be used in future experiments. Finally we summarize our conclusions in Section \ref{sec:conclusions}.

\section{Schwinger effect}
\label{sec:Schwinger}

The instability of the quantum vacuum in strong electric fields was first anticipated by Heisenberg and Euler through an effective action approach for constant backgrounds \cite{Heisenberg:1936nmg} (see Ref. \cite{Weisskopf:1936hya} for scalar QED). Schwinger later formalized this result using the modern language of QED by computing the imaginary part of the one-loop effective action to evaluate the vacuum-persistence amplitude \cite{Schwinger:1951nm},
\begin{equation}
|{}_{\mathrm{out}}\langle 0 | 0 \rangle_{\mathrm{in}}|^2 = \exp\left(-2\mathrm{Im}\Gamma \right)\,.
\end{equation}
Its deviation from unity signals  pair production from the vacuum. In 3+1 dimensions, the imaginary part of the effective action for spinor QED in a constant, uniform electric field reads\footnote{Throughout this work, we choose units such that $\hbar = c = k_B = 1$.}
\begin{equation}
\frac{2\mathrm{Im}\Gamma}{V\Delta t} = \frac{(eE)^2}{4\pi^3} \sum_{n=1}^\infty \frac{1}{n^2} \exp\left(-\frac{n\pi m^2}{eE}\right)\,,
\label{eq:Schwinger-fermion-rate}
\end{equation}
with $V\Delta t$ the spacetime volume, and $e$, $m$, $E$ the particle charge, mass, and the external field strength, respectively. This expression reveals the exponential suppression of the process for subcritical fields, and defines the natural scale $ E_{\mathrm{crit}} = \pi m^2/e$ as the threshold where non-perturbative pair production becomes significant (see Ref. \cite{Dunne:2004nc} for an historical perspective and Ref. \cite{Schubert:2024vft} for a review on both historical and technical aspects).\\

In what follows we adopt the (equivalent) mode–function viewpoint based on frequency mixing and Bogoliubov transformations.\footnote{See Refs. \cite{Lozanov:2018kpk,Domcke:2019qmm,Ferreiro:2018qdi,Beltran-Palau:2020hdr,Alvarez-Dominguez:2023ten,Alvarez-Dominguez:2023zsk} for different aspects of the Schwinger effect within this formalism,  and Ref. \cite{VicenteGarcia-Consuegra:2025lkh} for a stochastic generalization in the effective action formalism.  See also Ref. \cite{Taya:2020dco} for an exact WKB treatment of time-dependent electric fields. For a detailed discussion of quantum field states and their statistical properties in unstable-vacuum settings and critical potential steps, see Refs. \cite{Gavrilov:2014gca,Gavrilov:2016cfd,Gavrilov:2019vyi}.} By analyzing the frequency mixing of solutions to the Klein–Gordon and Dirac equations in time-dependent backgrounds, we can identify the  natural {\it in} and {\it out} mode bases and compute the coefficients that relate them. This provides direct access to the particle content of the theory and, crucially for our purposes, to the quantum correlations between the produced pairs.\\

We consider the case of a constant and uniform electric field along the $z$ direction ($\vec{E} = E\, \hat {e}_z$).\footnote{Despite the electric field being constant, particle creation still occurs: the time-dependence of the problem is encoded in the potential vector.} We treat two cases: (i) no magnetic field ($\vec{B}=\vec{0}\,$), and (ii)   anti-parallel magnetic field ($\vec B=-B\, \hat{e}_z$). We choose the gauge such that the vector potential, in the two cases, reads
\be
A_{\mu} = \begin{pmatrix}0\\0\\0\\Et\end{pmatrix} \qquad \mathrm{or}\qquad A_{\mu} = \begin{pmatrix}0\\0\\Bx\\Et\end{pmatrix}\, .
\ee
We restrict ourselves to the case $eE>0$ and $eB>0$. In what follows, we solve the Klein–Gordon and Dirac equations in momentum space, identify the corresponding {\it in} and {\it out} mode bases in each case, and compute the Bogoliubov coefficients that relate them. Readers already familiar with these techniques may wish to skip directly to subsection \ref{subsec:Bogoliubov}.

\subsection{Bosons}
Let us consider a massive charged scalar field $\phi$ governed by the Klein-Gordon equation in the presence of an electromagnetic background
\begin{equation}\label{eq:KG}
(D_{\mu} D^{\mu} + m^2)\phi = 0\,, \qquad \text{with} \qquad D_{\mu} = \partial_{\mu} + i e A_{\mu}\, . 
\end{equation}
The appropriate inner product in the space of solutions  between two field configurations $\phi_1$ and $\phi_2$ is given by the conserved symplectic form\footnote{The complex-conjugate field carries opposite charge and its (natural) KG product uses $D_\mu^*=\partial_\mu-i e A_\mu$. As a result, if $\phi$ has positive KG norm, $\phi^*$ has negative KG norm with respect to the conjugate product (see Ref. \cite{Alvarez-Dominguez:2025nce} for further details). }
\be
(\phi_1,\phi_2) = i\int \text{d}^3x \,(\phi_1^*\partial_t\phi_2 - \phi_2\partial_t\phi_1^*)\, .
\ee

\begin{itemize}
\item {\bf Case $\vec B=0$. } We start first with the pure electric field case. We decompose the scalar field along its momentum Fourier modes:
\be
\phi(\vec{x},t) = \int \text{d}^3k \,e^{i\vec{k}\cdot\vec{x}}\varphi_{\vec{k}}(u)\, .
\ee
and we introduce new variables
\be
u = \frac{k_z+eEt}{\sqrt{eE}} \qquad \mathrm{and}\qquad \lambda_k = \frac{k_x^2+k_y^2+m^2}{eE}\, .
\ee
In terms of the new variables, the equation  for the modes becomes
\be
 (\partial_u^2 + u^2 + \lambda_k)\varphi_{\vec{k}}(u) = 0\, \qquad \forall \vec{k}\,. 
\ee
The normalized solutions to this equation can be expressed in terms of parabolic cylinder functions $D_{\nu}(z)$
\be
d_{\pm 1}(\lambda_k,u) = \mathcal{N}_{\lambda_k} \, D_{-\frac{1+i\lambda_k}{2}}\left(\pm(1+i)u\right) \, , \qquad 
d_{\pm 2}(\lambda_k,u) = d_{-1}^*(\lambda_k,\pm u) \, ,
\ee
with
\be \label{eq:Nlambda}\mathcal{N}_{\lambda}=\frac{e^{-\frac{\lambda\pi}{8}}}{(2\pi)^{3/2}(2eE)^{\frac{1}{4}}} \, .\ee
These solutions can be grouped in pairs to form a complete basis. In particular, one can construct two orthonormal sets of solutions that are well-behaved in the asymptotic past ($u\to-\infty$) and future ($u\to+\infty$) corresponding respectively to the {\it in} and {\it out} bases:
\be
\mathcal{B}_{\mathrm{in}} = \Big\{ B_-^{\mathrm{in}}=d_{-1}(\lambda_k,u),B_+^{\mathrm{in}}=d_{2}(\lambda_k,u) \Big\} ~~~\mathrm{and}~~~ \mathcal{B}_{\mathrm{out}} = \Big\{ B_-^{\mathrm{out}}=d_{-2}(\lambda_k,u),B_+^{\mathrm{out}} = d_{1}(\lambda_k,u)\Big\}\, .
\ee
Note that due to $d_{-1}^*=d_2$ we have ${B^{\mathrm{in/out}}_+}^*=B^{\mathrm{in/out}}_-$.

\item {\bf Case $\vec{B}\parallel\vec{E}$.} 
We must solve again the Klein–Gordon equation \eqref{eq:KG}, but now consider the background with  electric and magnetic fields. In this case, we decompose the field by performing a Fourier transform in the transverse coordinates $\vec y=(y,z)$, and expand the $x$-dependence in terms of Hermite polynomials:
\be
\phi(x,\vec{y},t) = \int \text{d}^2k\,  e^{i\vec{k}\cdot\vec{y}}\sum_{l\in\mathbb{N}}e^{-\frac{\eta^2}{2}}H_l(\eta)\varphi_{\vec{k},l}(u)\, ,
\ee
where we have introduced the new variables
\begin{equation}
u = \frac{k_z+eEt}{\sqrt{eE}} ~~;~~ \eta = \frac{k_y+eBx}{\sqrt{eB}} ~~;~~ \lambda_l = \frac{m^2+(2l+1)eB}{eE}\, .
\end{equation}

The equation of motion for the modes reads
\be 
(\partial_u^2 + u^2 + \lambda_l)\varphi_{\vec{k},l}(u) = 0\, , \qquad \forall \vec{k}\,,\,~\forall l\, .
\ee
which has the same structure as the one encountered in the pure electric field case with the replacement  $\lambda_k \to \lambda_l$. Therefore,  we find the two same bases of solutions as in the pure electric field case (with $\lambda_k \to \lambda_l$)
\be
\mathcal{B}_{\mathrm{in}} = \Big\{ B_-^{\mathrm{in}},B_+^{\mathrm{in}} \Big\} ~~~\mathrm{and}~~~ \mathcal{B}_{\mathrm{out}} = \Big\{ B_-^{\mathrm{out}},B_+^{\mathrm{out}} \Big\}\, .
\ee

\end{itemize}

\subsection{Fermions}

A massive charged Dirac spinor $\psi$ obeys the equation of motion
\begin{equation}
(i\gamma^{\mu} D_{\mu} - m)\psi = 0 \,, \qquad \text{with}\qquad D_{\mu} = \partial_{\mu} + ieA_{\mu} \, .
\label{Dirac_eq_of_motion_fermions}
\end{equation}
We work in the Dirac representation of the gamma matrices. The inner product between two solutions 
of the Dirac equation is given by
\begin{equation}
(\psi_1, \psi_2) = \int \mathrm{d}^3x\, \psi_1^{\dagger}(\vec{x}, t)\psi_2(\vec{x}, t) \,.
\end{equation}
Unlike the bosonic case,  and for the sake of brevity, we do not treat here the $\vec B=0$ configuration separately, since the Bogoliubov coefficients can be obtained from the case with anti-parallel $\vec{E}$ and $\vec{B}$ fields in the limit $B\to 0$. More details can be found in Appendix \ref{ap:fermionsB0}. We proceed by decomposing the spinor field into Fourier modes along the $y$ and $z$ directions, and expanding its $x$-dependence in terms of Hermite polynomials:\footnote{The parameter $l$ of the Hermite polynomials is often called ``Landau level'' (for instance in \cite{Domcke:2019qmm}).} 
\be \label{eq:mode-expansion-fermions-BE}
\psi(x,\vec{y},t) = \int \text{d}^2\,k \,e^{i\vec{k}\cdot\vec{y}}\sum_{l\in\mathbb{N}}e^{-\frac{\eta^2}{2}}H_l(\eta)\left(\psi_1(\vec{k},l,u)\begin{pmatrix}1\\0\\1\\0\end{pmatrix}  + \psi_2(\vec{k},l,u)\begin{pmatrix}1\\0\\-1\\0\end{pmatrix} + \psi_3(\vec{k},l,u)\begin{pmatrix}0\\1\\0\\-1\end{pmatrix} + \psi_4(\vec{k},l,u)\begin{pmatrix}0\\1\\0\\1\end{pmatrix}\right)
\ee
where we introduced the new variables
\begin{equation}
u = \frac{k_z+eEt}{\sqrt{eE}} ~~~;~~~ \eta = \frac{k_y+eBx}{\sqrt{eB}} ~~~\mathrm{and}~~~ \lambda_l = \frac{m^2+2leB}{eE}\, .
\end{equation}
To find solutions, we solve the second order differential equation for an auxiliary spinor $F$, namely $(\slashed{D}^2+m^2)F(\vec p, l, u) = 0$, with $F(\vec p, l, u) =\sum_{i=1}^4  F_i(\vec p,l,u) \hat e_i$ and with $\hat e_i$ as in \eqref{eq:mode-expansion-fermions-BE}. From this ansatz we deduce a solution to the Dirac equation $\psi=(i\slashed{D}+m)F=\sum_{i=1}^4  \psi_i(\vec p,l,u) \hat e_i$.
After some algebra, one can find the two following orthonormal bases of solutions, corresponding to the {\it in} and {\it out} solutions that are well-behaved in the asymptotic past and future respectively: 
\be\label{eq:basisfermions01}
\mathcal{B}^{\mathrm{in}} = \Big\{ U_{\mathrm{in}}^L, U_{\mathrm{in}}^R, V_{\mathrm{in}}^L, V_{\mathrm{in}}^R\Big\} ~~~\mathrm{and}~~~ \mathcal{B}^{\mathrm{out}} = \Big\{ U_{\mathrm{out}}^L, U_{\mathrm{out}}^R, V_{\mathrm{out}}^L, V_{\mathrm{out}}^R\Big\}\, \, ,
\ee
where
\bea
U_{\mathrm{in/out}}^L(\vec{k},l,u) &=&  \sqrt{2(l+1)}\, \mathscr{N}_{l+1}\,L_{1/2}(\vec{k},l,u) \, ,\qquad \quad 
U_{\mathrm{in/out}}^R(\vec{k},l,u) =  \mathscr{N}_{l} \,\,R_{1/2}(\vec{k},l,u)\, , \\
V_{\mathrm{in/out}}^L(\vec{k},l,u) &=& \sqrt{\frac{4(l+1)}{\lambda_{l+1}}}\mathscr{N}_{l+1}\,\, L_{-2/-1}(\vec{k},l,u) \,, \qquad
V_{\mathrm{in/out}}^R(\vec{k},l,u) = \sqrt{\frac{2}{\lambda_l}}\, \mathscr{N}_{l}\,\,R_{-2/-1}(\vec{k},l,u)\, .
\eea
with the normalization constant given by 
\be
\mathscr{N}_l= \frac{1}{4\pi^{5/4}}\frac{e^{-\frac{\lambda_l\pi}{8}}}{\sqrt{eE2^l\, l!}}\, ,
\ee
and where the $L_i$ and $R_i$ functions are defined from the newly defined parabolic functions $S_i$ as: 
\be
L_{i}(\vec{k},l,u) = (i\slashed{D}+m)\,S_{i}(\lambda_{l+1},u)\begin{pmatrix}1\\0\\1\\0\end{pmatrix}
\, ; \quad  
R_{i}(\vec{k},l,u) = (i\slashed{D}+m)\,S_{i}(\lambda_l,u)\begin{pmatrix}0\\1\\0\\-1\end{pmatrix}.
\ee
with $i=\{\pm1,\pm2\}$ and 
\be
S_{\pm 1}(\lambda,u) = D_{-1+i\frac{\lambda}{2}}\left( \pm(-1+i)u\right) \, , \qquad 
S_{\pm 2}(\lambda,u) = D_{-i\frac{\lambda}{2}}\left(\pm(1+i)u\right)\, .
\ee
 Unfortunately, the $L$ and $R$ solutions are neither spin nor helicity nor chirality eigenstates.  However, in the massless limit ($m = 0$), they become eigenstates of chirality: the $L$ modes are left-handed ($\gamma^5 L = -L$), while the $R$ modes are right-handed ($\gamma^5 R = +R$). This property will prove useful when discussing the emergence of the chiral anomaly.

\subsection{Bogoliubov coefficients} \label{subsec:Bogoliubov}

 We now turn to the Bogoliubov transformations that relate the {\it in} and {\it out} solutions. We begin with the scalar case.
The set of {\it in} solutions 
$\{B^{\mathrm{in}}_-, B^{\mathrm{in}}_+\}$ can be written in terms of the {\it out} solutions 
$\{B^{\mathrm{out}}_-, B^{\mathrm{out}}_+\}$ through the Bogoliubov transformation:
\begin{equation}
\begin{dcases}
B_-^{\mathrm{in}}(\lambda,u) = \alpha_{\lambda}\, B_{-}^{\mathrm{out}}(\lambda,u) + \beta_{\lambda} \,B_+^{\mathrm{out}}(\lambda,u) \\
B_+^{\mathrm{in}}(\lambda,u) = \alpha'_{\lambda}\, B_+^{\mathrm{out}}(\lambda,u) + \beta'_{\lambda}\, B_-^{\mathrm{out}}(\lambda,u)
\end{dcases}\, ,
\end{equation}
where 
\begin{equation}
\begin{dcases}
\alpha_{\lambda} = \sqrt{2\pi}\frac{e^{-\lambda\pi/4-i\pi/4}}{\Gamma(\frac{1}{2}+\frac{i\lambda}{2})} \\
\beta_{\lambda} = ie^{-\lambda\pi/2} 
\end{dcases}
\qquad \mathrm{and}\qquad  
\begin{dcases}
\alpha'_{\lambda} = \alpha_{\lambda}^* \\
\beta'_{\lambda} = \beta_{\lambda}^*
\end{dcases}\, ,
\label{Bogoliubov_coeffs_schwinger_scalar}
\end{equation}
are the well-known Bogoliubov coefficients, that can be computed as inner products of the \textit{in} and \textit{out} solutions\cite{parker-toms}.\footnote{See Ref. \cite{Padmanabhan:1991uk} for a detailed computation of the Bogoliubov coefficients in the constant $E$-background case, and for interesting remarks on different gauge choices, the relationship with cosmological particle production, and connection to effective action methods.} Using the fact that $|\Gamma(\frac{1}{2}+\frac{i\lambda}{2})|^2 = \frac{\pi}{\cosh(\frac{\lambda\pi}{2})}$ one can verify that
\[
|\alpha_\lambda|^2-|\beta_\lambda|^2=1 \,,\qquad  |\alpha'_\lambda|^2-|\beta'_\lambda|^2=1 \qquad \mathrm{and}\qquad  \alpha_\lambda\beta_\lambda'^*-\alpha_\lambda'^*\beta_\lambda=0\,.
\]
which ensures that the operators in \eqref{Bogoliubov_transformation_scalar_no_B} and \eqref{Bogoliubov_transformation_scalar_with_B} satisfy the canonical commutation relations. The solutions indexed by ``$+$'' are positive energy solutions while the solutions indexed by ``$-$'' are negative energy solutions. When the field operator $\hat{\phi}$ is expanded in the {\it in} and {\it out} bases $\mathcal{B}_{\mathrm{in/out}}$, its coefficients with respect to the positive energy solutions in the {\it in} and {\it out} bases are identified with the annihilation operator at early and late times respectively and its coefficients with respect to the negative energy solutions are accordingly identified with the creation operators. This way it is easily found  that the {\it in} and {\it out} annihilation and creation operators are related by the Bogoliubov transformation. \\

 In the $B=0$ case, we have
\begin{equation}
\begin{pmatrix} \hat{a}_{\mathrm{out}}(\vec{k}) \\ \hat{b}_{\mathrm{out}}^{\dagger}(-\vec{k})\end{pmatrix} = \begin{pmatrix} \alpha_{\lambda_k} & \beta_{\lambda_k} \\ \beta'_{\lambda_k} & \alpha'_{\lambda_k}\end{pmatrix}\begin{pmatrix} \hat{a}_{\mathrm{in}}(\vec{k}) \\ \hat{b}_{\mathrm{in}}^{\dagger}(-\vec{k})\end{pmatrix}\, ,
\label{Bogoliubov_transformation_scalar_no_B}
\end{equation}
while in the $\vec{B}\parallel\vec{E}$ case, we have
\begin{equation}
\begin{pmatrix} \hat{a}_{\mathrm{out}}(\vec{k},l) \\ \hat{b}_{\mathrm{out}}^{\dagger}(-\vec{k},l)\end{pmatrix} = \begin{pmatrix} \alpha_{\lambda_l} & \beta_{\lambda_l} \\ \beta'_{\lambda_l} & \alpha'_{\lambda_l}\end{pmatrix}\begin{pmatrix} \hat{a}_{\mathrm{in}}(\vec{k},l) \\ \hat{b}_{\mathrm{in}}^{\dagger}(-\vec{k},l)\end{pmatrix}\, .
\label{Bogoliubov_transformation_scalar_with_B}
\end{equation}

Let us move now to the spinor case. Just as in the scalar case, the relationship between the {\it in} and {\it out} bases comes from the relationship between the different $S_i$ functions. In the $\vec{B}\parallel\vec{E}$ case, we find the following Bogoliubov transformations between the creation and annihilation operators at early and late times (in agreement with Refs. \cite{Nikishov:1969tt,Soldati:2011gi}) 

\begin{equation}
\label{Bogoliubov_transformation_spinor_with_B}
\begin{pmatrix} \hat{u}_{L,\mathrm{out}}(\vec{k},l) \\ \hat{v}_{L,\mathrm{out}}^{\dagger}(-\vec{k},l) \\ \hat{u}_{R,\mathrm{out}}(\vec{k},l) \\ \hat{v}_{R,\mathrm{out}}^{\dagger}(-\vec{k},l)\end{pmatrix} = \begin{pmatrix} \alpha^{(L)}_{l} & \beta^{(L)}_{l} &0&0 \\ \beta^{\prime (L)}_{l} & \alpha^{\prime (L)}_{l} &0&0 \\ 0&0& \alpha^{(R)}_{l} & \beta^{(R)}_{l} \\ 0&0& \beta^{\prime (R)}_{l} & \alpha^{\prime (R)}_{l} \end{pmatrix} \begin{pmatrix} \hat{u}_{L,\mathrm{in}}(\vec{k},l) \\ \hat{v}_{L,\mathrm{in}}^{\dagger}(-\vec{k},l) \\ \hat{u}_{R,\mathrm{in}}(\vec{k},l) \\ \hat{v}_{R,\mathrm{in}}^{\dagger}(-\vec{k},l)\end{pmatrix}\, ,
\end{equation}
where the Bogoliubov coefficients read (see, for example, Ref. \cite{Soldati:2011gi} for a detailed derivation)
\begin{equation} 
\begin{dcases} 
\alpha^{(L)}_{l} = \frac{\sqrt{\lambda_{l+1}\pi}}{\Gamma(1-i\frac{\lambda_{l+1}}{2})}e^{-\lambda_{l+1}\pi/4} \\
\beta^{(L)}_{l} = e^{-\lambda_{l+1}\pi/2}
\end{dcases}
~;\qquad 
\begin{dcases}
\alpha^{\prime (L)}_{l} = \alpha^{*(L)}_l \\
\beta^{\prime (L)}_{l} = -\beta^{*(L)}_l
\end{dcases}\, ,
\label{Bogoliubov_coeffs_schwinger_spinor_u}
\end{equation}
\begin{equation}
\begin{dcases}
\alpha^{(R)}_{l} = \frac{\sqrt{\lambda_{l}\pi}}{\Gamma(1-i\frac{\lambda_{l}}{2})}e^{-\lambda_{l}\pi/4} \\
\beta^{(R)}_{l} = e^{-\lambda_{l}\pi/2}
\end{dcases}
~;\qquad 
\begin{dcases}
\alpha^{\prime(R)}_{l} = \alpha^{*(R)}_{l} \\
\beta^{\prime(R)}_{l} = -\beta^{*(R)}_{l}
\end{dcases}\, .
\label{Bogoliubov_coeffs_schwinger_spinor_v}
\end{equation}
Using $|\Gamma(1-\frac{i\lambda}{2})|^2 = \frac{\lambda\pi}{2\sinh(\frac{\lambda\pi}{2})}$ we can verify that we have:
\[
|\alpha_{l}|^2 + |\beta_{l}|^2 = 1 \,,\qquad  |\alpha'_{l}|^2 + |\beta'_{l}|^2 = 1 \qquad \mathrm{and}\qquad  \alpha_{l}\beta^{\prime*}_{l}+\alpha_{l}^{\prime *}\beta_{l}=0\,.
\]
Note that everything happens as if our spinor $\hat{\psi}$ was composed of two scalars: $\hat{u}_L$ and $\hat{u}_R$, each of them evolving with its pair of Bogoliubov coefficients $(\alpha^{(L)},\beta^{(L)})$ and $(\alpha^{(R)},\beta^{(R)})$.  Note also that the Bogoliubov coefficients depend solely on the parameter $\lambda$, which enables us to easily switch from the $\vec{B}=0$ case where $\lambda = \lambda_k$ for both $R$ and $L$ fermions to the $\vec{B}\parallel \vec{E}$ case where $\lambda=\lambda_l$ for $R$ fermions and $\lambda=\lambda_{l+1}$ for $L$ fermions. \\

To connect this picture to the effective action language, we show in Appendix \ref{ap:fermionsB0} how to obtain the vacuum persistence amplitude from the Bogoliubov coefficients for the case $\vec B \parallel \vec E$.

\section{Quantum entanglement}
\label{sec:entanglement}

As we are interested in quantifying the amount of entanglement in the Schwinger effect, both for bosons and fermions, we provide a pedagogical presentation of the tools and formalism we leverage from quantum information theory. We begin by defining some of the most popular entanglement and correlations quantifiers we use in our calculations, more specifically, entanglement entropy, logarithmic negativity, and mutual information, and then we illustrate how to compute the entanglement in some specific cases of interest.

\subsection{Quantifying entanglement}

Consider a quantum state described by the density matrix $\hat{\rho}$, acting on some Hilbert space $\mH$, which admits a partition $\mH=\mH_A\otimes \mH_B$. We say the state $\hr$ is \textit{separable} if and only if it can be written as 
\begin{equation}
\hr=\sum_{i=1}p_i\,\hr_A^{(i)}\otimes \hr_B^{(i)}, \label{eq:separability}
\end{equation}
where $\hr_A^{(i)}$ and $\hr_B^{(i)}$ are density matrices acting on $\mH_A$ and $\mH_B$, respectively, and $0\leq p_i\leq 1,\; \forall\; i$. If a state $\hr$ cannot be written as in (\ref{eq:separability}), then it is said to be \textit{entangled}.\\

Quantifying the amount of entanglement in quantum states is a topic of central interest in quantum information theory and there exists a whole program of entanglement quantification with several measures \cite{Plenio:2006} being employed for this task. The most widely known, among them, is the von Neumann entropy of the reduced state, computed by $S_V=-\Tr_A[\hr_A\log_2\hr_A]$, with $\hr_A=\Tr_B\hr$, where $\Tr_B$ denotes the partial trace over subsystem $B$, while $\Tr_A$ in the entropy formula denotes the ordinary trace over the Hilbert space of subsystem $A$. For a pure bipartite state, one equivalently has $S_V(\hr_A)=S_V(\hr_B)$ (the result is independent on whether tracing out subsystem $A$ or $B$).  This measure, however, is limited to the cases where the total state $\hr$ is pure. This is rarely the case in real-life setups since pure states are associated to isolated systems. In realistic frameworks, systems are always exposed, to some degree, to noise resulting from interactions between the system of interest and its environment, i.e. the set of undesired degrees of freedom. \\

We are interested in incorporating the effect of noise, modelled as thermal fluctuations, in our study of entanglement in the Schwinger effect. To this end, we employ the \textit{logarithmic negativity} $L_N$ \cite{Plenio:2005}, which is based on the \textit{Peres-Horodecki} or \textit{positivity of the partial transposition} (PPT) criterion \cite{Peres:1996,Horodecki:1996} for bosonic systems. According to this criterion, a state $\hr$ is separable only if the partially transposed density operator $\hat{\tilde{\rho}}=\bm{I}_A\otimes \bm{T}_B(\hr)$ is positive semidefinite, i.e. $\hat{\tilde{\rho}} \geq 0$, where $\bm{T}_B$ is the operation of transposition with respect to the $B$-sector (the criterion is invariant under the choice of the sector with respect to which one transposes). From the latter, it follows that a state is entangled if there exists at least one negative eigenvalue of $\hat{\tilde{\rho}}$. The violation of the PPT criterion is, thus, a \textit{sufficient} condition for entanglement, but not a necessary one, in general.\footnote{It has been shown that the violation of the PPT criterion is both necessary and sufficient condition for systems with Hilbert space dimensions $2 \times 2$ and $2 \times 3$ \cite{Horodecki:1996}, as well as for Gaussian continuous variable systems in which one subsystem is made of a single degree of freedom or the state is isotropic \cite{Adesso:2006}.} However, we are interested in the scenario where each of the subsystems is made of a single degree of freedom (or mode), for which the PPT is both necessary and sufficient. \\

Finally, the logarithmic negativity is defined by
\begin{equation}
L_N=\log_2\left(\sum_{i}|\tilde{\lambda}_i|\right), \label{eq:LogNeg}
\end{equation}
where $\tilde{\lambda}_i$ are the eigenvalues of $\hat{\tilde{\rho}}$.\\

This PPT prescription does not directly apply to fermionic systems, where the associated operators satisfy anticommutation relations rather than commutation relations. In particular, it was shown \cite{Shapourian:2016cqu} that using logarithmic negativity for fermionic quantum states based on the PPT criterion discussed in the previous paragraph, which we may also refer to as \textit{bosonic PPT}, one fails to capture some of the fermionic quantum correlations. The criterion was extended for fermionic systems in \cite{Shapourian:2016cqu}, where the fermionic partial transposition was introduced, and according to which the fermionic logarithmic negativity was defined to quantify the amount of entanglement in fermionic states. Furthermore, it was shown in \cite{Shapourian:2018ozl} that it is an entanglement monotone under local operations and classical communication. The result of the fermionic partial transposition compared to the bosonic one is the introduction of phases in some of the density matrix elements (see section \ref{Quantum signal of the Schwinger effect} for the details  and Ref. \cite{Florio:2023mzk} for an example in 1+1.). Finally, to compute entanglement in the fermionic system, we make use of (\ref{eq:LogNeg}) but applied to the density matrix transformed under the fermionic partial transposition.

\subsection{Gaussian formalism of continuous variable systems} \label{sec:gaussiancontinuous}
Continuous variable systems are those whose observables possess a continuous spectrum, such as the harmonic oscillator. Upon quantization, the states of such systems are elements of an infinite-dimensional Hilbert space, and, hence, computing entanglement in such systems amounts to manipulating infinite-dimensional density matrices. Gaussian systems, however, such as those generated by quadratic Hamiltonians, enjoy a drastic simplification: they are fully characterized by their first and second moments, whose size depends on the number of degrees of freedom. We now briefly review the Gaussian formalism and how to employ it to compute entanglement. Given that we are interested in the mode of pairwise entanglement, we will restrict the discussion to the case of two degrees of freedom. A pedagogical reference for the Gaussian state formalism and entanglement in Gaussian states can be found in \cite{Serafini:book}.\\

Consider a system of two modes, for which one can associate the ladder operators $\{\ha_1,\had_1,\ha_2,\had_2\}$. From them, one can define the Hermitian ``position" and ``momentum" operators
\begin{equation}
\hx_J=\frac{\ha_J+\had_J}{\sqrt{2}}, \quad \hp_J=-i\frac{\ha_J-\had_J}{\sqrt{2}}, \quad J=1,2. \label{eq:xpa_tranformation}
\end{equation}
that satisfy the canonical commutation relations $[\hx_J,\hp_{J^\pr}]=i\delta_{JJ^\pr}$. One can further define the vector of operators $\bm{\hR}=(\hx_1,\hp_1,\hx_2,\hp_2)^\top$. By means of this vector, the commutation relations can be compactly written in the matrix form
\begin{equation}
[\bm{\hR},\bm{\hR}^\top]=i\bm{\Omega}, \quad \text{where} \quad \bm{\Omega}=\begin{pmatrix}
0&1&0&0\\
-1&0&0&0\\
0&0&0&1\\
0&0&-1&0
\end{pmatrix}.
\end{equation}
Any density operator $\hr$ can be uniquely reconstructed knowing all its moments. For a Gaussian state $\hr$, it suffices to know the first and second moments. Those are defined by
\begin{align}
\bm{\mu}&=\Tr\left[\bm{\hR}\bm{\hr}\right],\label{eq:mu}\\
\bm{\sigma}&=\Tr\left[\left\{\bm{\hR}-\bm{\mu},\bm{\hR}^\top-\bm{\mu}^\top\right\}\bm{\hr}\right]. \label{eq:sigma}
\end{align}
where $\left\{\cdot,\cdot\right\}$ is the anticommutator. In the definitions of the covariance matrix $\bm{\sigma}$, we subtracted $\bm{\mu}$ to remove any redundant information tied to first moments, and we took the anticommutator to remove the state-independent contribution of the commutator. The importance of realization in this formalism is that all the information about the infinitely dimensional $\bm{\rho}$ is uniquely encoded in the pair of finite-dimensional matrices $(\bm{\mu},\bm{\sigma})$. In particular, all information about correlations and entanglement are exclusively contained in $\bs$.\\ 

As a simple demonstration, we show how the mean number of quanta, the von Neumann entropy, the mutual information, and the logarithmic negativity can be efficiently computed for Gaussian states.\\

First, the average particle number contained in the state of the $J$-th degree of freedom is given by
\begin{equation}
\langle \hat{a}_J^\dagger \hat{a}_J \rangle=\frac{1}{4}\Tr\left[\bm{\sigma}_{\text{red},J}\right]+\frac{1}{2}\bm{\mu}_{\text{red},J}^\top \bm{\mu}_{\text{red},J}-\frac{1}{2}, \label{eq:number_of_quantum_from_sigma}
\end{equation}
where ``red" indicates the reduced subspace of the full $\bm{\mu}$ $\bm{\sigma}$, i.e. the quantities obtained by tracing out the rest of the degrees of freedom. Notice that (\ref{eq:number_of_quantum_from_sigma}) can be generalized for a system containing $N$ degrees of freedom by replacing the $1/2$ term at the end with $N/2$. The von Neumann entropy of the $J$-th mode is given by
\begin{equation} 
S_{V,J}=\frac{\nu_{J}+1}{2}\log_2\left(\frac{\nu_J+1}{2}\right)-\frac{\nu_J-1}{2}\log_2\left(\frac{\nu_J-1}{2}\right), \label{eq:von_Neumann_entropy_from_sigma}
\end{equation}
where $\nu_J$ is the positive-definite symplectic eigenvalue of $\bm{\sigma}$, i.e. the positive\footnote{The symplectic eigenvalues of an $2N\times 2N$ covariance matrix come in $N$ pairs of a positive and a negative eigenvalue, which are identical in absolute value.} eigenvalues of the matrix $i\bm{\Omega}\bm{\sigma}$. The latter can easily be extended to evaluate the entropy of a system comprised of $N$ degrees of freedom by computing the right-hand-side of (\ref{eq:von_Neumann_entropy_from_sigma}) for each of the $N$-positive-definite symplectic eigenvalues and then taking the sum of them.

To quantify the total number of correlations, both quantum and classical, one can evaluate the mutual information. This is readily computed utilizing the von Neumann entropy. For example, for the system of two degrees of freedom defined in this subsection, 
\begin{equation}
M_{\text{info}}=S_{V,1}+S_{V,2}-S_{V,{12}}, \label{eq:mutual_information}
\end{equation}
where $S_{V,1}$ and $S_{V,2}$ are the von Neumann entropies of each of the two subsystems, i.e. computed by the means of their reduced density matrices, while $S_{V,12}$ is the von Neumann entropy of the full system, and thus, is computed from the symplectic eigenvalues of the total covariance matrix.\\ 

Now, let us turn our attention to the logarithmic negativity. Simon extended the PPT criterion to two-mode continuous variable Gaussian systems \cite{Simon:2000}. The key step towards computing entanglement is the operation of the partial transposition. In terms of a $4\times 4$ covariance matrix $\bm{\sigma}$, the partial transposition reads \cite{Serafini:book}
\begin{equation}
\tilde{\bs}=\bm{T}\bs\bm{T}, \quad \text{where} \quad  \bm{T}=
\begin{pmatrix}
1&0&0&0\\
0&1&0&0\\
0&0&1&0\\
0&0&0&-1
\end{pmatrix}. \label{eq:sigma_PT}
\end{equation}
Recall that a signal of entanglement is the existence of a negative eigenvalue of the partially transposed density matrix. At the level of the covariance matrix, this is equivalent to the existence of at least one symplectic eigenvalue of $\tilde{\bs}$ smaller than 1, where the symplectic eigenvalues $\tilde{\nu}_i$ of $\tilde{\bs}$ are the eigenvalues of the matrix $i\bm{\Omega}\tilde{\bs}$. Finally, the logarithmic negativity measuring the entanglement in the two-mode Gaussian system under consideration is given by
\begin{equation}
L_N=\max\{0,-\log_2\tilde{\nu}_{\min}\}, \label{eq:LogNegCVG2}
\end{equation}
where $\tilde{\nu}_{\min}$ is the smallest positive, symplectic eigenvalue of $\tilde{\bs}$. Equation (\ref{eq:LogNegCVG2}) ensures that $L_N>0$ when $\tilde{\nu}_{\min}<1$ and zero otherwise. As previously emphasized, for a two-mode Gaussian state, such as in the bosonic Schwinger effect, the violation of the PPT criterion is both a necessary and sufficient condition. Hence, $L_N>0$ implies the existence of entanglement while $L_N=0$ guarantees its absence.\\

The Gaussian formalism can also be extended to the fermionic case, although it is not necessary to compute the entanglement quantifiers we study in this article. In Appendix \ref{app:gaussianforfermions} we provide a short operational approach to the Gaussian formalism for fermions, parallel to what we did in this section for bosons. In the following subsection we show, with an example, how to compute the most relevant objects for fermionic systems from the density matrix.

\subsection{Example: a two-level bipartite system}
\label{sec:twolevel}

Before diving into our results on the Schwinger effect, and for pedagogical purposes, let us first present the equivalent of a two-mode squeezing operation for fermions on general grounds. Consider the two-particle two-level system, initially prepared in the vacuum state $\ket{\psi}_{\tin}=\ket{0}_1\otimes\ket{0}_2$, with the two sectors interacting for a finite amount of time, inducing pair-creation. In the Heisenberg picture, this is described by the Bogoliubov transformation
\begin{equation}
\begin{aligned}
\ha_1^{\tout}&=\alpha \ha_1^{\tin}+\beta\ha_2^{\tin\dagger}\\
\ha_2^{\tout}&=\alpha \ha_2^{\tin}-\beta\ha_1^{\tin\dagger}, 
\end{aligned}
\label{eq:Bog_transf_fermions}
\end{equation}
with $\alpha, \beta \in \mathbb{C}$, $|\alpha|^2+|\beta|^2=1$, and the initial ladder operators $\hat{a}_i$ annihilate the initial vacuum state, i.e. $\ha_i\ket{0}_{i}=0$, $i=1,2$. In full generality, the initial state $\ket{0}_{\tin}$ can be expressed in the {\it out}-basis as
\begin{align}
\ket{0}_{\tin}=c_{00}\ket{0,0}_{\tout}+c_{01}\ket{0,1}_{\tout}+c_{10}\ket{1,0}_{\tout}+c_{11}\ket{1,1}_{\tout} \, .\label{eq:in_to_out}
\end{align}
In the {\it out}-basis, the vacuum state $\ket{0}_{\tout}$ is annihilated by $\ha_i^{\tout}$, $i=1,2$. Hence, solving (\ref{eq:Bog_transf_fermions}) in terms of the {\it in} ladder operators and plugging together with (\ref{eq:in_to_out}) into $\ha^{\tin}_i\ket{0}_{\tin}=0$ one finds $c_{01}=0=c_{10}$ and $c_{11}=c_{00}\beta/\alpha^*$. Choosing $c_{00}=\alpha^*$, one can write the state $\ket{0}_{\tin}$ in terms of the {\it out}-basis as\footnote{with the convention $|11\rangle = \hat{a}_2^{\dagger}\hat{a}_1^{\dagger}|0\rangle = -\hat{a}_1^{\dagger}\hat{a}_2^{\dagger}|0\rangle$}
\begin{equation}
\ket{0}_{\tin}=\alpha^*\ket{0,0}_{\tout}+\beta\ket{1,1}_{\tout} \label{eq:psi_out}\, .
\end{equation}

The corresponding density matrix is 
\begin{equation} \label{eq:rhoout_fermions}
\hr^{(\tout)}=\begin{pmatrix}
|\alpha|^2&0&0&\alpha^*\beta^*\\
0&0&0&0\\
0&0&0&0\\
\alpha\beta&0&0&|\beta|^2
\end{pmatrix}\, ,
\end{equation}
 where $\hat{\rho}^{(\mathrm{out})}$ can be  derived from $\hat{\rho}^{(\mathrm{in})}$ through the matrix implementation of eq. \eqref{eq:psi_out}, 
 \be \hat{\rho}^{(\mathrm{out})} = M\hat{\rho}^{(\mathrm{in})}M^{\dagger}\qquad  \text{where}\qquad  M=\begin{pmatrix}\alpha^*&0&0&-\beta^*\\0& 1&0&0\\0&0&1&0\\\beta&0&0&\alpha\end{pmatrix}\, .\ee
Using the definition of von Neumann entropy for the reduced state in the previous subsection, one can compute the amount of entanglement in state $\ket{\psi}_{\tout}$
\begin{align}
\nonumber
S_V&=-|\alpha|^2\log_2|\alpha|^2-|\beta|^2\log_2|\beta|^2\\
&=-(1-|\beta|^2)\log_2(1-|\beta|^2)-|\beta|^2\log_2|\beta|^2\, .
\end{align}
Note that the entanglement vanishes in either of the limits $|\beta|\rightarrow 0$, $|\beta|\rightarrow 1$, while it acquires its maximum value ($E_{V,\max}=1$) for $|\beta|=1/\sqrt{2}$, for which $\ket{\psi}_\tout$ (\ref{eq:psi_out}) becomes a Bell state and is, hence, maximally, entangled.\\

Let us now consider the scenario where the two particles are initially in a thermal state of temperature $T$. The two reduced states are not entangled and are given by 
\begin{equation}
\hr_i=\frac{\exp{\left(-\frac{\omega_i}{T}\hat{a}_i^\dagger \hat{a}_i\right)}}{\Tr\left[\exp{\left(-\frac{\omega_i}{T}\hat{a}_i^\dagger \hat{a}_i\right)}\right]}=\left(1-f_{i}\right)\ket{0}\bra{0}_i+f_{i}\ket{1}\bra{1}_i,
\end{equation}
where $\omega_i$ is the energy of the $i$-th particle and $f_{i}=\left(e^{\omega_i/T}+1\right)^{-1}$ is the mean occupation number of the $i$-th particle.
We consider $f_{1}=f_{2}=f$. The total state reads
\begin{equation} \label{eq:rhoin_fermions}
\hr^{(\tin)}=\begin{pmatrix}
(1-f)^2&0&0&0\\
0&f(1-f)&0&0\\
0&0&f(1-f)&0\\
0&0&0&f^2
\end{pmatrix}.
\end{equation}

The final density matrix reads 
\begin{equation} \label{eq:rhoout_fermions_thermal}
\hr^{(\tout)}=\begin{pmatrix}
(1-f)^2|\alpha|^2+f^2|\beta|^2&0&0&(1-2f)\alpha^*\beta^*\\
0&f(1-f)&0&0\\
0&0&f(1-f)&0\\
(1-2f)\alpha\beta&0&0&f^2|\alpha|^2+(1-f)^2|\beta|^2
\end{pmatrix}.
\end{equation}
Since the total state is mixed, we cannot use von Neumann entropy to quantify entanglement. Instead, one can use fermionic logarithmic negativity. We do so directly in section \ref{Quantum signal of the Schwinger effect} to compute entanglement in the Schwinger effect.

\subsection{Example: two-mode squeezing for different input states} \label{subsec:subsection-squeezing}
The underlying canonical transformation associated with bosonic particle creation in quantum field theory in curved spacetime, such as the Schwinger effect \cite{Schwinger:1951nm}, the Hawking \cite{Hawking:1974rv} and Unruh effects \cite{Unruh:1976}, and particle creation in cosmology \cite{Parker:1968mv, Grishchuk:1990}, is a two-mode squeezing. It is, thus, essential to review this transformation under the Gaussian state formalism and apply the tools laid down in the previous subsection to quantify entanglement for different input states of interest such as the vacuum, thermal, and squeezed states. All of them are non-displaced, and hence, we will consider $\bm{\mu}=\bm{0}$ for the remaining of the paper.\\

Consider two-modes of a field with their associate ladder operators $(\ha_1^{\tin},\ha_1^{\dagger\tin})$ and $(\ha_2^{\tin},\ha_2^{\dagger\tin})$. Working in the Heisenberg picture, the two-mode squeezing transformation reads
\begin{align}
\ha_1^{\tout}&=\alpha \ha_1^{\tin}+\beta\ha_2^{\dagger\tin}\\
\ha_2^{\tout}&=\alpha \ha_2^{\tin}+\beta\ha_1^{\dagger\tin},
\end{align}
where $\alpha, \beta\in \mathbb{C}$ are the Bogoliubov coefficients, satisfying $|\alpha|^2-|\beta|^2=1$. These are commonly expressed by means of the squeezing intensity $r$ and squeezing angle $\varphi$ (both being real parameters) as follows
\begin{equation}
\alpha=\cosh r, \quad \beta=e^{i\varphi}\sinh r.
\end{equation}
Using equation (\ref{eq:xpa_tranformation}), the definition of the vector $\bm{\hR}$, and the definition (\ref{eq:sigma}), it is a simple exercise to show that the two-mode squeezing process implies the following transformation of the covariance matrix
\begin{equation}
\bs_{\tout}=\bm{S}_{\text{sq}}\bs_{\tin}\bm{S}_{\text{sq}}^\top, \label{eq:sigma_transformation}
\end{equation}
where 
\begin{equation}
\bm{S}_{\text{sq}}=
\begin{pmatrix}
\cosh r&0&\cos \varphi \sinh r&\sin \varphi \sinh r\\
0&\cosh r&\sin \varphi \sinh r&-\cos \varphi \sinh r\\
\cos \varphi \sinh r&\sin \varphi \sinh r&\cosh r&0\\
\sin \varphi \sinh r&-\cos \varphi \sinh r&0& \cosh r
\end{pmatrix}, \label{eq:Ssq}
\end{equation}
where $\bs_{\tin}$ and $\bs_{\tout}$ are the initial and final covariance matrices, respectively. Note that $\bm{S}_{\text{sq}}$ is independent of the state it acts on. All the information about the initial state is encoded in $\bs_{\tin}$. \\

We now study the entanglement generated in the two-mode squeezing process for three different input states. We begin by considering that both modes are initially in their vacuum state. The corresponding initial covariance matrix is the identity $\bs_{\tin}=\bm{I}_4$. Substituting into (\ref{eq:sigma_transformation}), one obtains the evolved covariance matrix containing the correlations of the two-mode squeezed vacuum state. To compute entanglement, we plug the resulting covariance matrix into (\ref{eq:sigma_PT}) to obtain its partial transposition. Finally, we compute the symplectic eigenvalue of the partially transposed matrix and make use of (\ref{eq:LogNegCVG2}) to compute the amount of entanglement. The result is
\begin{equation}
L_N^{\text{vac}}=\frac{2}{\ln 2}r\approx 2.89\, r.
\end{equation}
Entanglement increases linearly with the squeezing strength. \\

As a next application, we consider both of the modes to be in a thermal state of temperature $T$. Thermal states are mixed states and can capture the effect of noise resulting from the interaction between the system of interest with its environment, which has a temperature $T$. The input covariance matrix reads $\bs_\tin=(2n+1)\bm{I}_4$, where $n$ is the mean number of quanta as a function of the frequency.\footnote{ Here we use $n$ and not $f$ since the system follows Bose-Einstein statistics.} Following the same steps for the computation of logarithmic negativity one finds
\begin{equation}
L_N^{\text{th}}=\max\left\{0,-\log_2\left[(2n+1)e^{-2r}\right]\right\}
\end{equation}
The latter is positive (thus resulting in an entangled final state) only when 
\begin{equation}
r>\frac{1}{2}\ln(2n+1). \label{eq:LogNeg_condition}
\end{equation}
The latter condition manifests the competition between the squeezing process that generates the entanglement and the thermal fluctuations degrading quantum correlations. For the final state to be entangled, the squeezing strength $r$ needs to dominate the noise intensity $n$. \\

The example of the thermal input highlights a crucial lesson: the non-classical signal (entanglement) of genuine quantum processes can be lost due to noisy inputs, hence masking the ``quantumness" of the process. Although one could overcome this challenge by increasing the two-mode squeezing parameter $r$, in several setups, there are physical constraints limiting the value of $r$. For example, in the Hawking effect, $r$ depends on the Hawking temperature which is very low for astrophysical black holes. In the Schwinger effect, $r$ depends on the strength of the electric field, which is also physically constrained (see discussion in Section  \ref{sec:experiment}). Hence, to revive the quantum output of the process, one may turn into modifying the input state. Introducing non-classicalities, i.e. features of states with strong departure from their classical field theory description, in the input can amplify the amount of entanglement produced during the two-mode squeezing process. One such non-classical state is the single-mode squeezed state.\footnote{Not to be confused with a two-mode squeezed state.} These are quantum states with uncertainty in one of their quadrature operators, i.e. position or momentum, below the uncertainty of the vacuum state, hence ``squeezed", whereas the conjugate quadrature exhibits higher-than the vacuum uncertainty, hence ``antisqueezed". Since noncommutativity is a characteristic of quantum mechanics, giving rise to an intrinsic uncertainty absent in classical mechanics, the antisqueezing of a quadrature, which increases this uncertainty, makes this nonclassical feature more prominent. It is also worth mentioning that single-mode squeezed states are not entangled; the squeezing is an operation on a single mode (degree of freedom). The idea of introducing non-classical inputs to enhance the amount of entanglement in particle creation processes was introduced in \cite{Agullo:2021vwj, Brady:2022ffk} where it was shown that the use of a single-mode squeezed state in the input can amplify the amount of entanglement produced by the two-mode squeezing process. In this paper we apply the same protocol for the Schwinger effect. The reason for using a single-mode squeezed state in the input instead of, for example other non-classical inputs, such as a two-mode squeezed state is to avoid artificially introducing entanglement in the system. \\

It is now interesting to review the interplay between noise and initial squeezing. To this end, let us consider the following single-mode squeezed thermal state 
\begin{equation}
\bs_\tin=(2n+1)\begin{pmatrix}
e^{2\xi_1}&0&0&0\\
0&e^{-2\xi_1}&0&0\\
0&0&e^{2\xi_2}&0\\
0&0&0&e^{-2\xi_2}
\end{pmatrix}, \label{eq:1SqTh_input}
\end{equation}
where $\xi_1$ and $\xi_2$ are the squeezing amplitudes of each mode. We have chosen the corresponding squeezing phases such that the squeezing happens in the momentum sector of each of the modes. Note that the later state is generated by acting $(\bm{S}_{\xi_1}\oplus \bm{S}_{\xi_2})\bs_{\text{th}}(\bm{S}_{\xi_1}^\top\oplus \bm{S}_{\xi_2}^\top)$, where $\bs_{\text{th}}=(2n+1)\bm{I}_4$ is the thermal state and 
\begin{equation} \label{eq:squeezing-initial-matrix}
\bm{S}_{\xi_i}=
\begin{pmatrix}
\cosh{\xi_i}-\cos{\theta_i}\sinh{\xi_i} & -\sin{\theta_i}\sinh{\xi_i}\\
-\sin{\theta_i}\sinh{\xi_i} & \cosh{\xi_i}+\cos{\theta_i}\sinh{\xi_i}
\end{pmatrix}, \quad i=1,2 
\end{equation}
is the single-mode squeezing symplectic matrix with amplitude $\xi_i$ and phase $\theta_i$. The latter determines the direction of the squeezing on the $\hat{x}_i-\hat{p}_i$ plane, and plays no role in the discussion of the entanglement. Hence, we choose $\theta_1=\theta_2=\pi$, for convenience, resulting in (\ref{eq:1SqTh_input}). Note that the single-mode squeezing (\ref{eq:squeezing-initial-matrix}) acts locally in the space of modes and, hence, it does not result in mode interactions, and consequently, it cannot produce entanglement. Hence, the state (\ref{eq:1SqTh_input}) is separable.  

Repeating the same procedure as before, one can obtain the evolved covariance matrix, its partial transposition with its symplectic eigenvalue, and, finally, the logarithmic negativity. Since the final expression is rather cumbersome, we plotted it below to illustrate the interplay of the several parameters.  For simplicity, we consider a symmetric single-mode squeezing, i.e. $\xi_1=\xi_2=\xi$.  We then study how the curve $L_N(n)$ depends on the input squeezing strength $\xi$, keeping fixed the two-mode squeezing strength value to $r=1$.
\begin{figure}[h!]
\centering
{\includegraphics[width=0.45\textwidth]{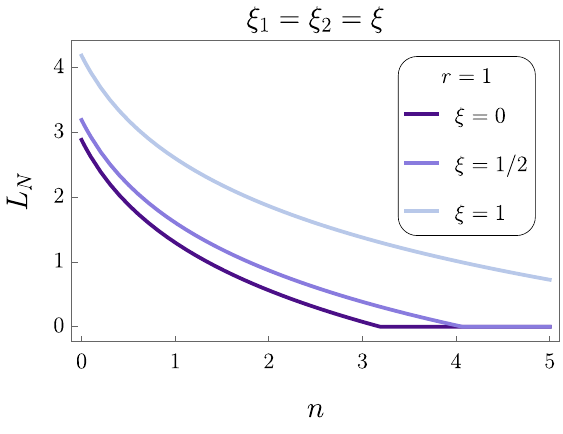}}
\caption{Logarithmic Negativity vs the noise intensity parameter $n$ for the symmetric single-mode squeezed thermal state, with $\xi_1=\xi_2=\xi$, after we evolve it with the two-mode squeezing transformation. For the plot, we use the two-mode squeezing strength value $r=1$ and we study different values of the input squeezing strength $\xi$.}
\label{fig:LogNeg_single_mode_thermal_state}
\end{figure}

Plot \ref{fig:LogNeg_single_mode_thermal_state} confirms the main point of the discussion in this section: entanglement decreases monotonically with $n$ and increases with $\xi$. In particular, notice that there is a cutoff value of $n$ after which entanglement vanishes. For the case $\xi=0$, this value is given by (\ref{eq:LogNeg_condition}). Introducing initial squeezing, i.e. $\xi\neq0$, shifts this cutoff to larger values of $n$. Hence, the entanglement produced during the two-mode squeezing gains robustness against thermal fluctuations. Let us also remark that in the absence of the two-mode squeezing process, i.e. $r=0$, there is not entanglement generated. This is crucial as it emphasizes that the initial squeezing $\xi$ does not generate entanglement, it rather amplifies it. The use of single-mode squeezed states at the input provides a promising avenue for experimental detection of the quantum features of particle creation processes in the laboratory. We use this protocol below to amplify the amount of entanglement in the bosonic Schwinger effect.

\section{Main results}
\label{sec:mainresults}

In this section, we present our main quantitative results using the language and tools developed in the previous sections. We present first the scalar case and then the spinor case, analyzing how statistics (i.e., bosonic or fermionic) affects both the number of created particles and the pattern of quantum correlations.

\subsection{Bosons}
We consider a system that is initially at thermal equilibrium at a temperature $T$ (eventually letting $T\to 0$ to study the vacuum state). Since we want to study bipartite correlations between a particle mode and its conjugate antiparticle mode, we describe them jointly as a two-mode sector $(I\bar I)$, where $\bar I$ denotes the mode with opposite momentum (and the same Landau/Hermite index when $\vec E\parallel\vec B$). The label $I$ collects the quantum numbers of the mode, namely
\be
I=\vec{k} \quad \text { when } \vec{B}=0, \quad I=(\vec{k}, l) \quad \text { when } \vec{E} \parallel \vec{B} .
\ee

The initial state is described by the thermal density matrix
\begin{equation}
\hat{\rho}^{(\mathrm{in})} = \bigotimes_{I}\hat{\rho}^{(\mathrm{in})}_{I\bar I}\, , \qquad  \hat{\rho}^{(\mathrm{in})}_{I\bar I}=  \frac{1}{Z_{I\bar I}} \sum_{n,\bar{n}=0}^{+\infty} e^{-\omega_I(n+\bar{n})/T}|n,\bar{n}\rangle_{I\bar I}^{(\mathrm{in})}\langle n,\bar{n}|_{I\bar I}^{(\mathrm{in})}\, .
\end{equation}
Here $\ket{n,\bar n}_{I \bar I}^{(\mathrm{in})}$ denotes a Fock state with $n$ particles of momentum $\vec k$ and Hermite parameter $l$ (or momentum $\vec k$ for $\vec B=0$), and $\bar n$ antiparticles with momentum $-\vec k$ and the same index $l$ (or momentum $-\vec k$ for $\vec B=0$). $Z_{I\bar I}$ is the normalized partition function (such that $\mathrm{Tr}(\hat{\rho}^{(\mathrm{in})}_{I\bar{I}})=1$), and $\omega_I$ is the one-particle energy 
\begin{equation} \label{eq:frequencybosons}
\omega_I^2 = m^2 + k_z^2 + k_{\perp}^2 ~~~\mathrm{where}~ k_{\perp}^2 = \begin{dcases}
k_x^2 + k_y^2 ~~\mathrm{if}~B=0 \\
(2l+1)eB ~~\mathrm{if}~B\neq 0
\end{dcases}\, .
\end{equation}

A thermal state is Gaussian. We can therefore use the formalism developed in Sec.~\ref{sec:gaussiancontinuous} and characterize it by its first and second moments. For each two-mode sector we have
\be \label{eq:instatebosons}
\bm{\mu}_{I\bar I}^{(\mathrm{in})}=\bm{0}\, , \qquad \text{and} \qquad \bm{\sigma}_{I\bar I}^{(\mathrm{in})}=(1+2n_I)\bm{I}_4\, ,\qquad \text{with}\qquad n_I=\frac{1}{e^{\omega_I/T}-1}\, .
\ee

\subsubsection{Number of created particles}

This initial state contains an average number of $\langle \hat{N}^{(\mathrm{in})}\rangle$ non-entangled particles described by the Bose-Einstein statistics. For a given mode $I$ (and its conjugate partner)  
\begin{equation}
\langle \hat{N}^{(\mathrm{in})}_{I}\rangle =n_I\,,\qquad \langle \hat{N}^{(\mathrm{in})}_{\bar I}\rangle =n_{\bar I}\equiv n_I\,.
\end{equation}
Therefore, for the two-mode system we are analyzing we find
\be
\langle \hat{N}^{(\mathrm{in})}_{I\bar I}\rangle=n_I+n_{\bar I}=2n_I\,.
\ee
This result can be obtained from the Gaussian formalism, that directly gives us the master formula to compute the particle number. In this case, since $N=2$, we find
\begin{equation}
\langle \hat N_{I\bar I}\rangle=\frac{1}{4}\,\Tr\,\bm\sigma_{I\bar I}+\frac{1}{2}\,\bm\mu_{I\bar I}^\top \, \bm\mu_{I\bar I}-1\,.
\label{eq:master_N_pair}
\end{equation}
For the {\it in} state, $\bm \mu$ and $\bm\sigma$ are given in \eqref{eq:instatebosons}.\\

To obtain the particle number for the {\it out} state, we shall now transform the initial density matrix with the Bogoliubov transformations given in \ref{subsec:Bogoliubov}. As these transformations do not mix the different modes, we can consider a fixed given pair $(I\bar I)$ and all the formulas will be equally applicable on all other modes. In terms of the Gaussian formalism, it is enough to apply the Bogoliubov transformation directly to the first and second moments 
\be \label{eq:bogosigmaandmuscalars}
\bm\mu^{(\mathrm{out})}_{I\bar I}=\bm S\,\bm\mu^{(\mathrm{in})}_{I\bar I}\,, \qquad \bm\sigma^{(\mathrm{out})}_{I\bar I}=\bm S\,\bm\sigma^{(\mathrm{in})}_{I\bar I}\bm S^\top ,
\ee
where $\bm S$ is the two-mode squeezer matrix given in eq. \eqref{eq:Ssq} with $r$ and $\varphi$ fixed by the Bogoliubov coefficients we computed in Section \ref{sec:Schwinger}  $(\alpha_I\equiv\alpha_{\lambda_I},\beta_I\equiv\beta_{\lambda_I})$. More explicitly\footnote{The parametrization of the two-mode squeezer assumes $\alpha_I$ is real, while the $\alpha_I$ we computed for the Schwinger effect is not. This is not a problem, since we can always perform a rotation that changes the phase of the Bogoliubov coefficients. This transformation does not affect particle number or entanglement \cite{Birrell:1982ix}.} 
\be
\cosh r= |\alpha_I|\, ,\qquad \sinh r = |\beta_I|\, ,\qquad  \varphi = \arg(\beta_I\, \alpha_I^*) \,.
\ee
After the Bogoliubov transformation \eqref{eq:bogosigmaandmuscalars}, we can use formula \eqref{eq:master_N_pair} for the {\it out} state.  The average number of particles and antiparticles reads 
\begin{equation}
\langle \hat{N}_I^{(\tout)} \rangle=\frac{1}{2}\langle \hat{N}_{I \bar I}^{(\text{out})}\rangle =  |\beta_I|^2 + \frac{|\alpha_I|^2+|\beta_I|^2} {e^{\omega_I/T}-1} = n_I + |\beta_I|^2\left(1+2n_I\right)=\frac{1}{2}((1+2n_I)\cosh 2 r -1)\, ,
\end{equation}
We note that for bosons we have thermal enhancement. That is, the {\it out} number of particles is not increased by $|\beta|^2$ but by $(1+2n_I)|\beta|^2$. It is  convenient to explicitly define the average difference 
\begin{equation} 
\Delta N_I =\frac{1}{2}\left(\langle \hat{N}_{I \bar I}^{(\text{out})}\rangle- \langle \hat{N}_{I \bar I}^{(\text{in})}\rangle\right) = |\beta_I|^2\frac{e^{\omega_I/T}+1}{e^{\omega_I/T}-1}\, .
\label{number_of_created_particles_scalar_schwinger_with_T}
\end{equation}
Using \eqref{eq:frequencybosons} and \eqref{Bogoliubov_coeffs_schwinger_scalar} we can write this expression in terms of electric and magnetic fields. The average number of created particles with momentum $(k_z,k_\perp)$ thus reads 
\begin{equation} \label{eq:differenceN-bosons}
\Delta N(k_z,k_\perp) = \exp\left(-\pi\frac{m^2+k_{\perp}^2}{eE}\right)\coth\left(\frac{\sqrt{m^2+k_z^2+k_{\perp}^2}}{2T}\right)\, ,
\end{equation}
 where $k^2_{\perp}=k_x^2+k_y^2$ if $B=0$ and $k^2_{\perp}=(2l+1)eB$ if $B\neq 0$. For example, if we want to look at the mode with the lowest momenta, we  set $k_z=k_\perp=0$ for the pure electric case and $k_z=l=0$ if $B\neq0$, which implies $k_\perp^2=eB$. \\

\begin{figure}[h]
\centering
{\includegraphics[width=0.45\textwidth]{"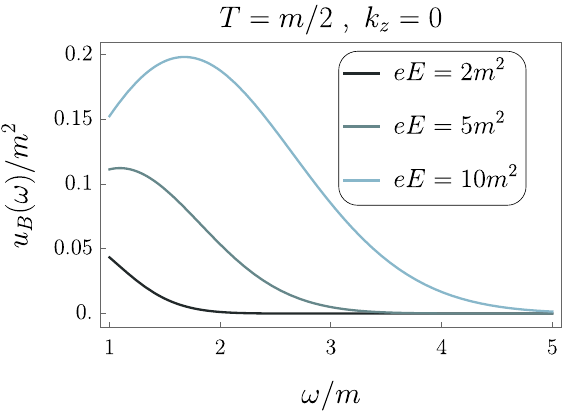"}}
\caption{Energy density $u_B(\omega)$ as expressed in (\ref{eq_energy_density_bosons_no_kz}).}
\label{energy_density_per_mode_noB_nom_no_kz_fixedT_bosons}
\end{figure}

Rather than in the number of created particles, we might be interested in their energy spectrum. For a given mode $I$ (with frequency $\omega_I$) the average energy increment per mode reads
\begin{equation}
\langle \Delta \mathcal{E}_I\rangle = \omega_I \Delta N_I=\omega_I |\beta_I|^2\frac{e^{\omega_I/T}+1}{e^{\omega_I/T}-1}=\omega_I|\beta_I|^2\coth\,\Big(\frac{\omega_I}{2T}\Big)\, .
\end{equation}
Note that $|\beta_I|^2$ is independent of $k_z$, which is because we are working with a constant electric field, so modes with any value of $k_z$ will eventually be produced. This means that the energy spectrum behaves, at high $k_z$, as $n(\omega) \sim \omega^2$ (without exponential suppression in $k_z$), which gives a divergent result for the average energy density, since we are considering particles that are created for an infinitely long time.\footnote{For a time-dependent electric field we would obtain a $k_z$-dependent $\beta$ coefficient that vanishes at $|k_z|\to+\infty$. This is, for instance, the case for a pulsed electric field (see for example Ref. \cite{Li_2017} or Ref. \cite{Beltran-Palau:2019bvz} in 1+1). In the literature, for the case with constant $E$, a cutoff off in the longitudinal momentum $k_z$, namely $0<k_z<eE\Delta t$ is used. This cutoff has been verified numerically in   Ref. \cite{Domcke:2019qmm}.}
To get rid of this problem, we can consider, for example, only particles with no momentum along the $z$ direction $k_z=0$ and find (in the $B=0$ case)
\begin{equation}
\frac{\Delta \mathcal{E}}{L^2} = \int_{k_z=0} \frac{\text{d}k_x \text{d}k_y}{(2\pi)^2}\omega_{\vec{k}}|\beta_{\lambda_{\vec{k}}}|^2\coth\left(\frac{\omega_{\vec{k}}}{2T}\right) \equiv \int_m^{+\infty}\text{d}\omega \, u_B(\omega)\, ,
\end{equation}
where $u_B(\omega)$ is given by
\begin{equation}
u_B(\omega) = \frac{\omega^2}{2\pi}e^{-\pi\frac{\omega^2}{eE}}\coth\left(\frac{\omega}{2T}\right)\, ,
\label{eq_energy_density_bosons_no_kz}
\end{equation}
and is represented in Figure \ref{energy_density_per_mode_noB_nom_no_kz_fixedT_bosons}.

\subsubsection{Critical field for particle creation}

At fixed temperature, inverting eq. \eqref{eq:differenceN-bosons} gives  the electric field needed to produce an average of $\Delta N$ particles with momentum $(k_z,k_\perp)$,
\be
E_{\Delta N}^{(T)}=\frac{\pi m^2}{e}\left(\frac{1+k_\perp^2/m^2}{ \ln \coth \frac{\omega}{2T}  - \ln \Delta N} \right)\, .
\ee
In particular, if we fix $\Delta N=e^{-1}$  and work in the low-momentum regime (we assume here $\vec B=0$ so we can fix $k_z=k_\perp=0$), we find, at zero temperature,
\be
E_{\text{crit}}\equiv E_{e^{-1}}^{(0)}= \pi \frac{ m^2}{e}\, ,
\ee
in agreement with the standard result, and its generalisation at finite temperature 
\be \label{eq:EcritT-particlenumber}
E_{\text{crit}}^{(T)}=\frac{\pi m^2}{e}\left(\frac{1}{1 + \ln \coth\frac{m}{2T}}\right) \underset{T\gg m}{\approx} \frac{\pi m^2}{e \ln(\frac{2T}{m})} \, .
\ee
 This shows that, for bosons in the high temperature regime, the critical electric field decreases logarithmically with the temperature (in other words, it is easier to create a pair). Conversely, for fixed electric field, inverting eq. \eqref{eq:differenceN-bosons} gives 
\begin{equation}
T_{\Delta N} = \frac{\sqrt{m^2+k_z^2+k_{\perp}^2}}{2\tanh^{-1}\left(\frac{1}{\Delta N}\exp\left(-\pi\frac{m^2+k_{\perp}^2}{eE}\right)\right)}\, .
\end{equation}
Setting, for instance, $\Delta N=1$ gives the temperature $T_1$ required to observe in average 1 pair of created particles for a given electric field $E$, or the minimum temperature such that we still have reasonably high chances of observing a pair. As an example, for very low energy modes $\omega_k\simeq m$ and for very weak electric fields $eE\ll m^2$ we have $T_1 \simeq \frac{m}{2}e^{\pi m^2/eE}$.

\subsubsection{Quantum signal of the Schwinger effect}

Now, we turn our attention to the quantum aspects of the Schwinger effect. In particular, employing logarithmic negativity as a suitable entanglement measure, we quantify the amount of entanglement produced and study how it depends on several model parameters.\\

\begin{figure}[h!]
\centering
\subfloat[]{
\includegraphics[height=0.35\textwidth]{"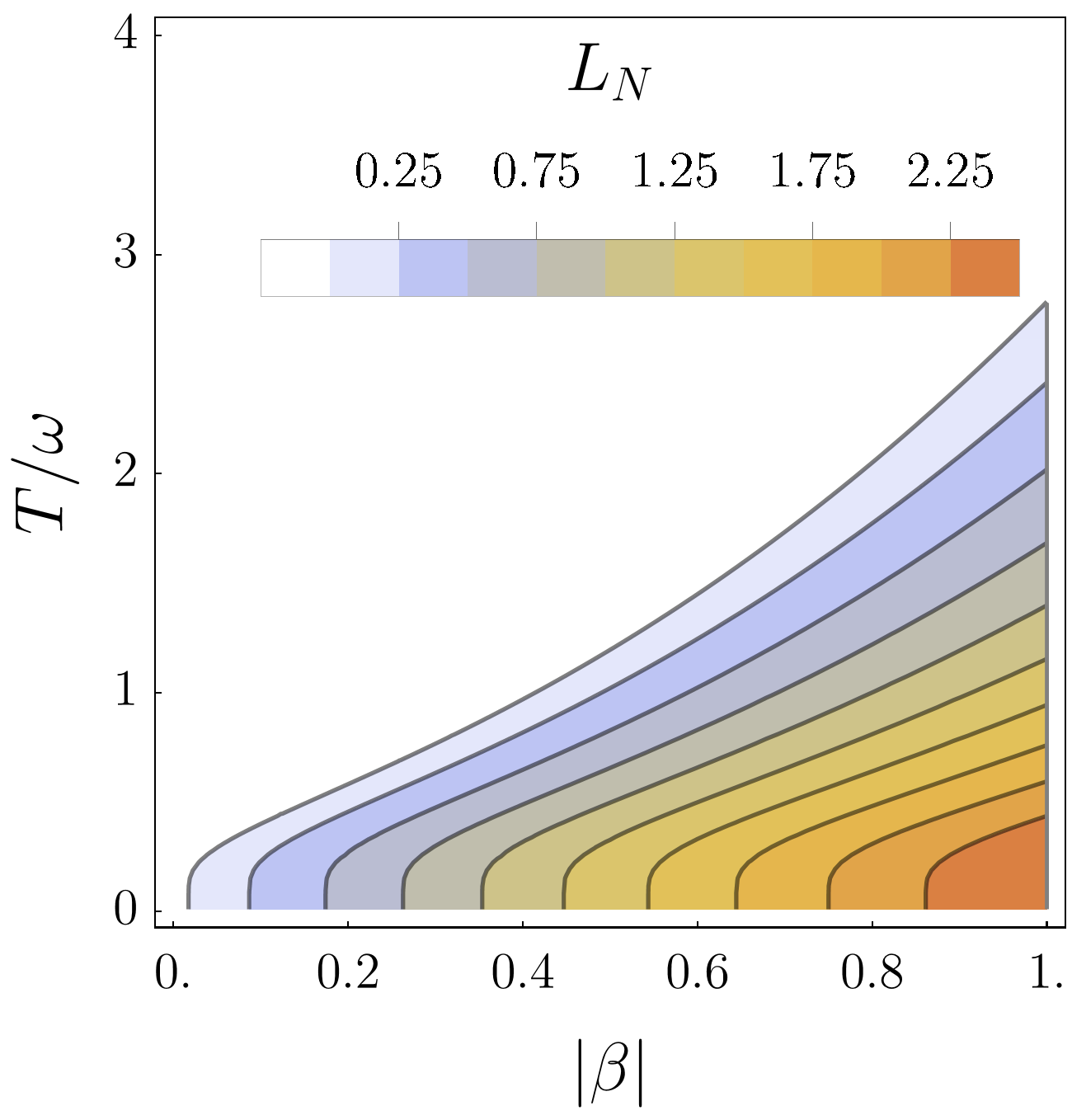"}}
\hspace{50pt}
\subfloat[]{\includegraphics[height=0.35\textwidth]{"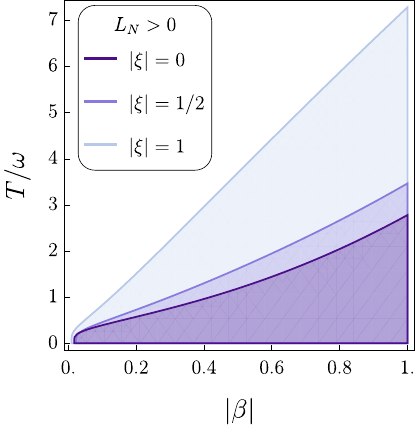"}}
\caption{(a) Phase diagram-like representation of the logarithmic negativity of bosons. We clearly see that there exists a temperature $T_c/\omega$ above which there can't be any entanglement because $|\beta|$ can't be higher than $1$. (b) Same for different values of squeezing, the shaded areas correspond to regions where $L_N>0$.}
\label{fig_phase_diagrams_bosons_with_T}
\end{figure}

To compute the logarithmic negativity, we follow the steps given in Section \ref{sec:gaussiancontinuous}. Recall the formula 
$L_N=\max\{0,-\log_2\tilde{\nu}_{\min}\},$ where $\tilde{\nu}_{\min}$ is the smallest positive symplectic eigenvalue of the partially transposed $4\times 4$ covariance matrix $ \tilde{ \bm \sigma}$ defined in eq. \eqref{eq:sigma_PT}. We want to focus on the {\it out} state, so we use the covariance matrix given in eq. \eqref{eq:bogosigmaandmuscalars}. It is not difficult to find
\bea
\tilde{\nu}_{\min}=e^{-2r}(1+2n_I)
=(1+2n_I)\big(|\beta_I|-\sqrt{1+|\beta_I|^2}\big)^2\, .
\eea
where we have used $0<|\beta_I|<1$. Therefore, in terms of $T$ and $\beta_I$, we directly find
\begin{equation}
L_N = \frac{1}{\ln(2)} \max\left(0,
-\ln\left(\frac{e^{\omega_I/T}+1}{e^{\omega_I/T}-1}(\sqrt{1+|\beta_I|^2}-|\beta_I|)^2\right)\right)\, .
\label{logneg_from_bogo_with_T_bosons}
\end{equation}
In Figure \ref{fig_phase_diagrams_bosons_with_T}, we  present in a phase-like diagram the logarithmic negativity as a function of $\beta_I$ and $T$.

It is easy to see that the higher $|\beta_I|$ is, the more entangled the system, and the higher the temperature, the lower the entanglement. Furthermore,  for a given pair $(I \bar I)$ there is entanglement (in the sense $L_N>0$) if and only if $\tilde \nu _{\text{min}}<1$, or, equivalently
\begin{equation} \label{eq:condition-critical-entanglement}
|\beta_I|^2 > \frac{1}{e^{2\omega_I/T}-1}\, , \qquad \text{i.e., }\qquad  \frac{T}{\omega_I} < \frac{T_c}{\omega_I} = \frac{2}{\ln\left(1+|\beta_I|^{-2}\right)}\, .
\end{equation}
 We also note that the higher $\omega_I$ is, the higher the critical temperature is. For a given pair $(I\bar I)$, the critical temperature to have entanglement in terms of electric and magnetic fields reads
\begin{equation} \label{eq:Tc-bosons}
T_c = \frac{2\sqrt{m^2+k_z^2+k_{\perp}^2}}{\ln\left(1+\exp\left(\pi\frac{m^2+k_\perp^2}{eE}\right)\right)}\, .
\end{equation}
For a given mode,  $T_c$ increases monotonously with the electric field $E$, reaching a constant value when $E\to \infty$, namely
\be
T_{c}\underset{E \to\infty}{=}\frac{2}{\ln(2)}\omega= T_{\text{max}}\, ,
\ee
which corresponds to the limit $|\beta_I|^2\to 1$.

\subsubsection{Critical field for entanglement}
We can do a similar analysis aiming to find a critical electric field for entanglement. The critical value is given by the condition \eqref{eq:condition-critical-entanglement}. 
To have an entangled pair $(I \bar I)$ with frequency $\omega_I$ at a given temperature $T<T_{\text{max}}$, the electric field should be higher than a critical value  $E>E_{\text{entang}}$, where 
\be\label{eq:Ecrit-entanglement-bosons}
e E_{\text{entang}}^{(I)}=\pi \frac{m^2 + k_\perp^2}{\ln(e^{2\omega/T}-1)}\, .
\ee
Some comments need to be made. First, we recall that the critical field is mode-dependent. Furthermore,  at zero temperature $T\to 0$, $eE_{\text{entang}}\to 0$ independently of $\omega_I$. In other words, any nonzero field produces a nonzero squeezing and the pair is entangled. On the other hand, for $T\to T_{\text{max}}(\omega)$, $eE_{\text{entang}}\to \infty$. For a fixed $k_{\perp}$ and fixed $T$, there exists a value of $|k_z|$ above which there is entanglement ($L_N>0$). The reason is that the entanglement threshold ($T_{\text{max}}$) depends on $\omega_I$ while $|\beta_I|$ is independent of $k_z$.\footnote{We note that this property is characteristic of the constant electric field case. For example, for a Sauter pulse, $\beta_I$ depends on $k_z$.}\\

\begin{figure}[h]
\centering
\includegraphics[width=0.45\textwidth]{"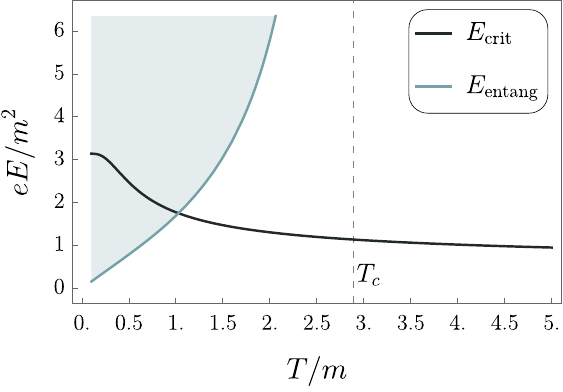"}
\caption{Critical electric fields as expressed in eqs. \eqref{eq:EcritT-particlenumber} and \eqref{eq_E_for_entanglement_bosons}. The shaded area corresponds to the region where the system is entangled. Above $T=T_c$, no entanglement is possible.}
\label{fig_critical_electric_fields_bosons}
\end{figure}

If one wishes to create entanglement then the minimum value required for the intensity of the electric field is given by $E_{\mathrm{entang}}^{k_{\perp}=0}$:
\begin{equation}
E_{\mathrm{entang}} = \frac{\pi m^2}{e\ln(e^{2m/T}-1)} \, ,
\label{eq_E_for_entanglement_bosons}
\end{equation}
which should be compared to the critical electric field for particle creation \eqref{eq:EcritT-particlenumber}, as in Figure \ref{fig_critical_electric_fields_bosons}.
We see that at low temperatures $T\ll m$ it is easier to create entanglement than to create an average of $\Delta N=e^{-1}$ particles ($E_{\mathrm{crit}}>E_{\mathrm{entang}}$), while at high temperatures $T\to T_{\mathrm{max}}$, it becomes the opposite.

\subsubsection{Initial squeezed state} \label{sec:initialsqueezed}

\begin{figure}[h]
\centering
\includegraphics[width=0.45\textwidth]{"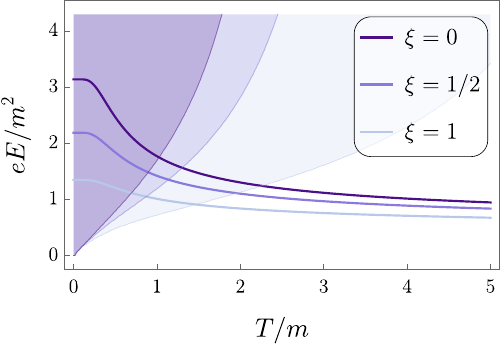"}
\caption{Critical electric fields $E_{\text{crit}}$ and $E_{\text{entang}}$ as expressed in \eqref{eq:Ecrit_squeezed_bosons}  and \eqref{eq:conditionEentangxi}  in the $\vec{k}=0$ mode. The shaded area correspond to the regions where the system is entangled and are evaluated numerically.  We see, as already explained in subsection \ref{subsec:subsection-squeezing} that the squeezing increases the entanglement and thus decreases the minimum electric field for entanglement $E_{\mathrm{entang}}$. The squeezing also stimulates the pair production, thus reducing the minimum electric field for particle creation $E_{\mathrm{crit}}$.}
\label{fig_critical_electric_fields_bosons_squeezed}
\end{figure}

We now present a strategy to amplify the amount of entanglement produced in the bosonic Schwinger effect. In particular, we consider an initial single-mode squeezed thermal state at temperature $T$ as introduced in Section \ref{subsec:subsection-squeezing}. In this case, the first and second moments are
\be 
\bm{\mu}_{I\bar I}^{(\mathrm{in})}= 0 \, , \qquad \text{and} \qquad \bm{\sigma}_{I\bar I}^{(\mathrm{in})}=(1+2n_I)\begin{pmatrix}
\bm S_{\xi}\bm S^\top_\xi & 0\\
0 & \bm S_{\xi}\bm S^{\top}_\xi
\end{pmatrix}\,,
\ee
where $\bm S_{\xi}$ is given in eq. \eqref{eq:squeezing-initial-matrix} with $\theta=\pi$ and where we have used the same squeezing parameter $\xi>0$ for the two modes of the pair. The out state is given, as before, by the transformation in eq. \eqref{eq:bogosigmaandmuscalars}.
In this case, the average number of created particles with momentum $(k_z,k_\perp)$ reads
\be
\Delta N_I= \sinh^2r(2n_I+1)\cosh(2\xi)\equiv|\beta_I|^2 \coth\left(\frac{\omega}{2T}\right) \cosh(2\xi)\, .
\ee
This result must be compared with the thermal one given in eq. \eqref{number_of_created_particles_scalar_schwinger_with_T}. At fixed $T$ we can compute the electric field required to observe an average of $\Delta N=e^{-1}$ particles with momentum $(k_z,k_\perp)$. For low-momentum pairs (we choose $\vec B=0$ so that $k_z=k_\perp=0$ is the lowest energy mode), we obtain the critical field
\bea\label{eq:Ecrit_squeezed_bosons}
E_{\text{crit}}^{(T;\xi)}&=& \frac{\pi m^2}{e}\left(\frac{1}{1 + \ln \coth\frac{m}{2T}+\ln(\cosh(2\xi))}\right) \\
&\underset{T\to0}{=}& \frac{\pi m^2}{e}\frac{1}{1+\ln \cosh(2\xi)}\underset{\xi\to\infty}{=}\frac{\pi m^2}{e} \frac{1}{2\xi} \, .
\eea
An initial squeezed state thus reduces the critical electric field required to observe particle production. This is due to the stimulated character of the process. To obtain the critical electric field to have entanglement at a given temperature $T<T_{\text{max}}$, we have to look at the logarithmic negativity 
$
L_N=\frac{1}{\ln(2)}\max\{0,-\ln\tilde{\nu}_{\min}\}$. The critical  electric field to have entanglement $E_{\text{entang}}^{(T;\xi)}$ is given by the implicit equation
\be \label{eq:conditionEentangxi}
\tilde \nu_{\text{min}}(E_{\text{entang}};T, \xi)=1\, .
\ee
In Figure \ref{fig_critical_electric_fields_bosons_squeezed} we represent the critical electric fields for relevant particle production and to have entanglement as a function of the temperature for different values of $\xi$.

\subsubsection{Mutual information}
To quantify the total amount of correlations in the Schwinger effect in each mode pair, we compute the mutual information shared between a given mode $I$ and its partner $\bar I$ (particle/anti-particle pair). The mutual information is defined in \eqref{eq:mutual_information}. To compute $S_{V,12}$ we use the out covariance matrix of the total system, while to compute $S_{V,1}$ and $S_{V,2}$ we use the reduced covariance matrix of the sub-system corresponding to mode $I$ (describing particles), i.e the upper-left $2\times 2$ block of the full covariance matrix, or the lower-right $2\times 2$ block for the mode $\bar{I}$ (describing antiparticles)
\be
\bm {\sigma}_I^{(\text{in})}= (1+2n_I)\bm I_2=\bm {\sigma}_{\bar I}^{(\text{in})}\, , \qquad \bm {\sigma}_I^{(\text{out})}= (1+2n_I) \cosh(2r) \bm I_2 = \bm {\sigma}_{\bar I}^{(\text{out})}\qquad \text{and} \qquad \bm \Omega_2= \begin{pmatrix}
0&1\\
-1&0
\end{pmatrix}\, .
\ee
where we again used $n_I=n_{\bar I}$. We use eq. \eqref{eq:von_Neumann_entropy_from_sigma} to compute the von Neumann entropies. To this end, we need to find the (positive) eigenvalue of the symplectic covariance matrix $i \bm\Omega \bm\sigma$ and insert it in \eqref{eq:von_Neumann_entropy_from_sigma}. The relevant symplectic value for the full system $(12)\equiv(I\bar I)$ reads
\be
\nu_{I\bar I}=1+2n_I\, ,\ee
while for the sub-systems 1 ($I$) and 2 ($\bar I$) we find  
\be
\nu_I=\nu_{\bar I}=\left(2|\beta_I|^2+1\right) \frac{e^{\omega_I / T}+1}{e^{\omega_I / T}-1}=\cosh(2r)(1+2n_I)\,.
\ee
Therefore, using eqs. \eqref{eq:von_Neumann_entropy_from_sigma} and \eqref{eq:mutual_information}, the mutual information of the {\it out} state reads  (in terms of $|\beta_I|$ and $T$)
\be
\begin{aligned}\label{eq:Minfo-bosons}
M_{\mathrm{info}} =& \frac{2}{\ln(2)}\left( |\beta_I|^2\frac{e^{\omega_I/T}+1}{e^{\omega_I/T}-1} \ln\left(\frac{e^{\omega_I/T}+|\beta_I|^2(1+e^{\omega_I/T})}{1+|\beta_I|^2(1+e^{\omega_I/T})}\right)\right.
\\
&+ \frac{1}{1-e^{-\omega_I/T}}\ln\left(1+|\beta_I|^2(1+e^{-\omega_I/T})\right) 
\left. -\frac{1}{e^{\omega_I/T}-1}\ln\left( 1+|\beta_I|^2(1+e^{\omega_I/T})\right) \right)\, .
\end{aligned}
\ee
For bosons, the mutual information decreases smoothly with noise (i.e., with $T$) and increases with $|\beta_I|^2$, or, equivalently, with the electric field $E$. Contrary to the logarithmic negativity, the mutual information is a continuous function that does not have a cut-off.\\

\subsection{Fermions}

Following the previous section, we now consider a Dirac field initially in thermal equilibrium at temperature $T$ (we may later take $T\to 0$ to recover the vacuum). It is described by the density matrix
\begin{equation}
\hat{\rho}^{(\mathrm{in})} = \bigotimes_I \left(\hat{\rho}_{I\bar I}^{(\text{in};R)}\otimes \hat{\rho}_{I\bar I}^{(\text{in};L)}\right)\, ,
\end{equation}
where 
\begin{equation}
\hat{\rho}_{I\bar I}^{(\text{in},i)} = \frac{1}{Z_{I\bar I}^{(i)}}\sum_{n,\bar{n}=0}^{1}e^{-\omega_{I}^{(i)}(n+\bar{n})/T}|n,\bar{n}\rangle_{I\bar I}^{(\mathrm{in};i)}\langle n,\bar{n}|_{I\bar I}^{(\mathrm{in};i)}\, ,
\end{equation}
with $i=(R,L)$, and where $I=(\vec k,l)$ when $\vec E\parallel\vec B$ and $I=\vec k$ when $\vec B=0$. The initial density matrix $\hat \rho_{I \bar I}^{(\text{in},i)}$ is given in eq. \eqref{eq:rhoin_fermions}. As for the scalar case, the frequency $\omega_I^{(i)}$ of a mode $I$ depends on the magnetic field as follows:
\begin{equation} \label{eq:frequencyfermions}
\omega_I^{(i)} = \sqrt{m^2+k_z^2+k_{\perp}^2} 
\qquad \mathrm{where}\qquad
k_{\perp}^2 = \begin{dcases}
k_x^2+k_y^2 \qquad\qquad \qquad\,\,\,\, \mathrm{if}~ B=0 \\
2leB\qquad\quad\,\mathrm{for~type~} R~\mathrm{if}~B\neq 0\\
2(l+1)eB ~~~\mathrm{for~type~}L ~\mathrm{if}~B\neq 0\\ 
\end{dcases}\,\,\,\, .
\end{equation}
 In what follows, we study the number of created particles as well as the entanglement profile of this system. From now on, we omit the superindex $^{(i)}$. All the results presented below are valid for the two types of modes ($L$ and $R$).

\subsubsection{Number of created particles}

Fermions at an initial temperature $T$ are described by the Fermi-Dirac statistics 
\be
\langle \hat{N}^{(\mathrm{in})}_{I}\rangle =f_I=\langle \hat{N}^{(\mathrm{in})}_{\bar I}\rangle\, ,\qquad \text{with}\qquad f_I=\frac{1}{e^{\omega_I/T}+1} \, .
\ee
Or, equivalently, for the pair particle/anti-particle, 
\be
\langle \hat{N}^{(\mathrm{in})}_{I\bar I}\rangle =2f_I\, .
\ee
 The {\it in} density matrix in the {\it out}-basis  $\{|00\rangle, |01\rangle, |10\rangle, |11\rangle\}$ reads (as before, we label it as $\hr^{(\tout)}$)
\be  \label{eq:out_density_matrix_fermions}
\hr^{(\tout)}_{I \bar I}=\begin{pmatrix}
(1-f_I)^2|\alpha_I|^2+f_I^2|\beta_I|^2&0&0&(1-2f_I)\, \alpha^*_I \beta^*_I\\
0&f_I(1-f_I)&0&0\\
0&0&f_I(1-f_I)&0\\
(1-2f_I)\, \alpha_I \beta_I &0&0&(1-f_I)^2|\beta_I|^2+f_I^2|\alpha_I|^2
\end{pmatrix}\, . 
\ee
It can be also useful to give the density matrix of the reduced systems in the {\it out}-basis (with only particles or only antiparticles), it reads
\be \label{eq:reducedrhoout}
\hat \rho_I^{\text{(out)}}=\text{Tr}_{\bar I}[\hr^{(\tout)}_{I \bar I}]= \begin{pmatrix}
(1-f_I)^2|\alpha_I|^2+f_I^2|\beta_I|^2 +f_I (1-f_I)&0\\
0&(1-f_I)^2|\beta_I|^2+f_I^2|\alpha_I|^2 +f_I (1-f_I)
\end{pmatrix}\, ,
\ee
and $\hat\rho_{\bar I}^{\text{(out)}}=\hat\rho_{I}^{\text{(out)}}$. The average number of particles (and of antiparticles) of a given type ($R$ or $L$) and in a given mode $I$ ($\bar I$) is given by 
\begin{equation}
\langle \hat{N}^{(\mathrm{out})}_I\rangle = |\beta_I|^2 + \frac{1-2|\beta_I|^2}{1+e^{\omega_I/T}} = f_I + |\beta_I|^2\left(1-2f_I\right)\, .
\label{number_of_created_particles_spinor_schwinger_with_T}
\end{equation} 
In this case we see that, contrary to the scalar case, less than $|\beta_I|^2$ particles have been created. This is logical because the number of particles in a given mode is limited to 1 due to the Pauli exclusion principle. The particles already present initially in the thermal state thus decrease the number of particles that can be created.
It is convenient to define explicitly the difference $\Delta N_I=\langle \hat{N}^{(\mathrm{out})}_I\rangle-\langle \hat{N}^{(\mathrm{in})}_I\rangle$. In terms of electric and magnetic fields, the number of created particles reads (for both $L$ and $R$ type)
\begin{equation} \label{eq:deltaNfermions}
\Delta N(k_z,k_\perp) = \exp\left(-\pi\frac{m^2+k_{\perp}^2}{eE}\right)\tanh\left(\frac{\sqrt{m^2+k_z^2+k_{\perp}^2}}{2T}\right)\, .
\end{equation}
We recall that for the $B=0$ case $k_\perp^{2}=k_x^2+k_y^2$, while for the case with a magnetic field $k_\perp^{2}=2leB$ for the $R$-type particles and $k_\perp^{2}=2(l+1)eB$ for the $L$-type particles. So this formula is valid for all cases and particle types.\\

\begin{figure}[h]
\centering
{\includegraphics[width=0.45\textwidth]{"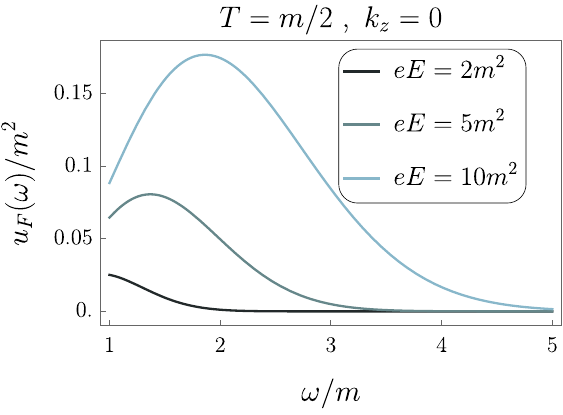"}}
\caption{Energy density $u_F(\omega)$ as expressed in (\ref{eq_energy_density_fermions_no_kz}). Contrarily to the scalar case, the higher the temperature, the less energy is produced from the electric field. At a fixed value of $T$ however, both bosonic and fermionic spectra have the same shape, with a smooth peak that appears at high electric field.}
\label{energy_density_per_mode_noB_fixed_T_no_kz_fermions}
\end{figure}

For completeness, we can also compute the average created energy in mode $I$ of frequency $\omega_I$.  The average energy increment per mode reads
\begin{equation}
\langle \Delta \mathcal{E}_I\rangle = \omega_I \Delta N_I=\omega_I |\beta_I|^2\frac{e^{\omega_I/T}-1}{e^{\omega_I/T}+1} =\omega_I|\beta_I|^2 \tanh\left(\frac{\omega_I}{2T}\right)\,
\end{equation}
and as in the scalar case, integrating $\Delta\mathcal{E}_I$ over the modes $I$ gives a spectrum that diverges at high energies due to the independence of $\beta_I$ in $k_z$. Just as we did in the scalar case we can simply integrate over $k_x$ and $k_y$ at $k_z=0$ (in the $\vec{B}=0$ case) to find:
\begin{equation}
\frac{\Delta \mathcal{E}}{L^2} = \int_{k_z=0} \frac{\text{d}k_x \text{d}k_y}{(2\pi)^2}\omega_{\vec{k}}|\beta_{\lambda_{\vec{k}}}|^2\tanh\left(\frac{\omega_{\vec{k}}}{2T}\right) = \int_m^{+\infty}\text{d}\omega \, u_F(\omega)\, ,
\end{equation}
where $u_F(\omega)$ is given by:
\begin{equation}
u_F(\omega) = \frac{\omega^2}{2\pi}e^{-\pi\frac{\omega^2}{eE}}\tanh\left(\frac{\omega}{2T}\right)\, .
\label{eq_energy_density_fermions_no_kz}
\end{equation}
and is represented in figure \ref{energy_density_per_mode_noB_fixed_T_no_kz_fermions}.

\subsubsection{Critical field for particle creation}

As in the scalar case, we can compute the electric field needed to produce an average of $\Delta N$ particles with momentum $(k_z,k_\perp)$. Inverting \eqref{eq:deltaNfermions} we find 
\be
E_{\Delta N}^{(T)}=\frac{\pi m^2}{e}\left(\frac{1+k_\perp^2/m^2}{ \ln \tanh \frac{\omega}{2T}  - \ln \Delta N} \right)\, .
\ee
Furthermore, we can define the critical electric field for particle creation as the minimum electric field such that $\Delta N > e^{-1}$. We thus have $e^{-\pi m^2/eE}\tanh\left(\frac{m}{2T}\right)>e^{-1}$, which gives
\begin{equation}\label{eq:Ecrit_fermions_with_T}
E_{\mathrm{crit}}^{(T)} = \frac{\pi m^2}{e(1+\ln\tanh(\frac{m}{2T}))}\, .
\end{equation}
As in the scalar case, we recover the well-known $E_{\mathrm{crit}}=\pi m^2/e$ in the $T\to 0$ limit, but contrary to that case, the critical electric field here increases with the temperature, reaching an infinite value $E_{\mathrm{crit}}\to+\infty$ when $T\to T_{\mathrm{max}} = \frac{m}{2\text{arctanh}(e^{-1})}$.\\

For a fixed electric field, eq. \eqref{eq:deltaNfermions} can be also inverted to obtain the required temperature to observe an average of $\Delta N$ particles (with $\Delta N$ between zero and 1) with a given momentum (and/or Hermite number) $I$, 
\begin{equation}
T_{ \Delta N} = \frac{\sqrt{m^2+k_z^2+k_{\perp}^2}}{2\coth^{-1}\left(\frac{1}{\Delta N}\exp\left(-\pi\frac{m^2+k_{\perp}^2}{eE}\right)\right)}\, .
\end{equation}

\subsubsection{Quantum signal of the Schwinger effect} \label{Quantum signal of the Schwinger effect}

We proceed with logarithmic negativity to evaluate the amount of entanglement produced. For fermions, we use directly the defining equation given in \eqref{eq:LogNeg}, since the density matrix is finite, namely
$L_N=\log_2(\sum_{i}|\tilde{\lambda}_i|)$
where ${|}\tilde{\lambda}_i{|}$ are the eigenvalues of  the absolute value of the partially  ``time reversed" density matrix $\sqrt{\hat{\tilde{\rho}}\hat{\tilde{\rho}}^{\dagger}}$. We focus directly on the {\it out}-basis. In our case, $\rho_{I \bar I }^{(\text{out})}$ is given in \eqref{eq:out_density_matrix_fermions}, and $\hat{\tilde \rho}_{I \bar I }$ is obtained as follows: if the components of the density matrix in the Fock basis $\{|ij\rangle \langle k l|\}_{i,j,k,l=0,1}$ are $\rho_{i j k l}$, then the components of $\hat{\tilde \rho}_{I \bar I}$ are $(-1)^{\phi_{ijkl}}\rho_{kjil}$  where $\phi_{ijkl}$ is a phase that depends on the occupation numbers $i,j,k,l$ as follows:
\begin{equation}
\phi_{ijkl} = \frac{(i+k) ~[\mathrm{mod}~2]}{2} + (i+k)(j+l)\,.
\end{equation} 
That is, we exchange the upper-right and lower-left $2\times 2$ blocks in \eqref{eq:out_density_matrix_fermions}  and give them an additional $\pm i$ factor and get
\be
\hat{\tilde\rho}^{(\text{out})} =\begin{pmatrix}
(1-f_I)^2|\alpha_I|^2+f_I^2|\beta_I|^2 & 0 & 0 & 0 \\
0 & f_I(1-f_I) & -i(1-2f_I)\, \alpha_I \beta_I & 0 \\
0 & -i(1-2f_I)\, \alpha_I^*\beta_I^* & f_I(1-f_I) & 0\\
0 & 0 & 0 & (1-f_I)^2|\beta_I|^2+f_I^2|\alpha_I|^2
\end{pmatrix}\, .
\ee
 Its absolute value is
\be
|\hat{\tilde\rho}^{(\text{out})}| \equiv \sqrt{\hat{\tilde\rho}^{(\text{out})}\hat{\tilde\rho}^{(\text{out})\dagger}} =\begin{pmatrix}
(1-f_I)^2|\alpha_I|^2+f_I^2|\beta_I|^2&0&0&0\\
0&s_I&0&0\\
0&0 &s_I&0\\
0&0&0&(1-f_I)^2|\beta_I|^2+f_I^2|\alpha_I|^2
\end{pmatrix}\, .
\ee
with $s_I=\sqrt{f_I^2(f_I-1)^2+|\alpha_I|^2 |\beta_I|^2 (1-2f_I)^2}$.\\

\begin{figure}[h!]
\centering
{\includegraphics[height=0.35\textwidth]{"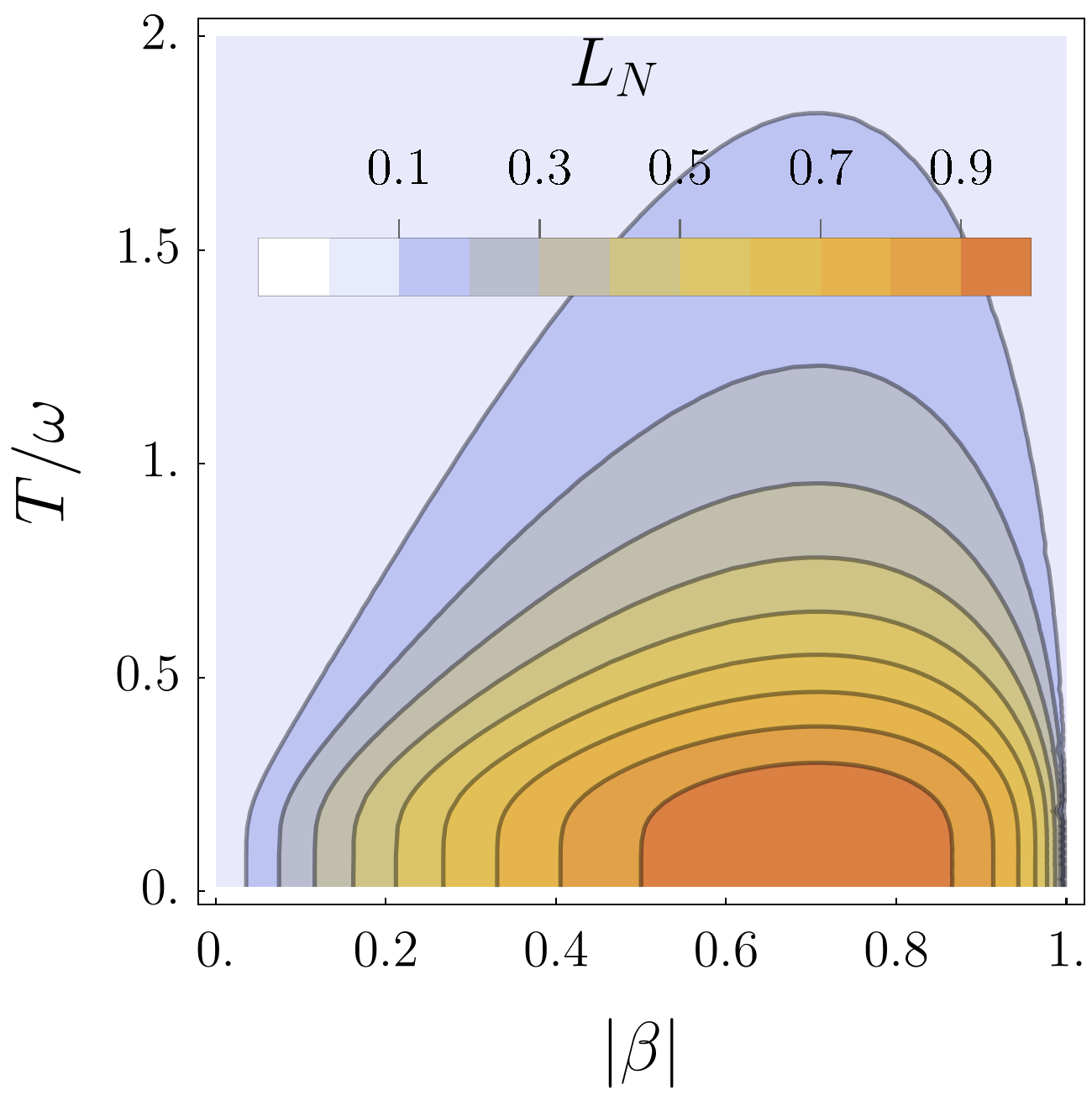"}}
\caption{ Phase-like diagram representation of the logarithmic negativity of fermions.  Contrarily to the bosonic case, as long as $|\beta|>0$ there exists no temperature above which the logarithmic negativity vanishes. }
\label{fig_phase_diagrams_fermions_with_T}
\end{figure}

The eigenvalues of $|\hat{\tilde\rho}^{(\text{out})}|$ can then be summed to yield the (fermionic) logarithmic negativity:
\be
L_N = \log_2(\text{Tr}|\hat{\tilde\rho}^{(\text{out})}|) = \frac{1}{\ln(2)}\ln\left(f^2_I + (1-f_I)^2 +2\sqrt{f_I^2(1-f_I)^2+(1-2f_I)^2 |\alpha_I|^2|\beta_I|^2}\right)\, .
\ee
More explicitly, in terms of $\omega_I$, $T$ and $|\beta_I|^2$,
\be
L_N = \frac{1}{\ln(2)}\ln\left(1+e^{2\omega_I/T} + 2\sqrt{e^{2\omega_I/T}+(e^{2\omega_I/T}-1)^2|\beta_I|^2(1-|\beta_I|^2)}\right) - \frac{2}{\ln(2)}\ln(1+e^{\omega_I/T})\, .
\ee
In figure \ref{fig_phase_diagrams_fermions_with_T}, we plot $L_N$ as a function of $\omega$ and $T$ in a phase-like diagram. \\

 Contrary  to the bosonic case, the logarithmic negativity for fermions does not display a temperature cut-off. 
As discussed in Section \ref{sec:entanglement}, this behaviour relies on using the fermionic definition of logarithmic negativity based on partial time-reversal, which provides a well-behaved entanglement quantifier for fermionic systems \cite{Shapourian:2016cqu}. 
By contrast, if one applies the usual bosonic partial transpose to the same fermionic density matrix, one generally obtains a different quantity that can display, for example, spurious critical temperatures, reflecting the fact that the bosonic prescription is not a good entanglement measure for fermionic systems. In what follows, we will study the general behaviour of $L_N$ and some limiting cases.\\

First, we note that, at fixed $|\beta_I|$, the logarithmic negativity decreases as $T$ increases. In particular, for high temperatures we find that the logarithmic negativity decreases quadratically with the temperature, namely 
\be
L_N = \frac{(1-|\beta_I|^2)|\beta_I|^2}{\ln(2)}\frac{\omega^2}{T^2} + \mathcal{O}\left(\frac{\omega^4}{T^4}\right)\,.
\ee
On the other hand, the logarithmic negativity at fixed $|\beta_I|$ reaches its maximum at $T=0$ (or $f_I=0$)
\be \label{eq:LNT=0}
L_N^{(0)}=\frac{\ln(1+2|\beta_I|\sqrt{1-|\beta_I|^2})}{\ln(2)}\, .
\ee

 At fixed $T$ (or fixed $f_I$), we find that the logarithmic negativity has a maximum at $|\beta_I|^2=1/2$, 
\be \label{eq:maximumLNf}
L_N^{\text{(max)}}=\frac{\ln\left(2+4f_I(f_I-1)\right)}{\ln(2)}=\frac{1}{\ln(2)}\ln \left(1+\tanh^2\left(\frac{\omega_I}{2 T}\right)\right)\, .
\ee
Note that, $L_{N}^{(\text{max})}$ reaches its maximum value at $T= 0$, namely $L_N^{\text{(max;0)}}=1$. This is  expected, since  entanglement is bounded by the finite local Hilbert-space dimension \cite{Shapourian:2018ozl}.

In addition, we also note that $L_N$ goes to zero for $|\beta_I|^2\to 0$ and for $|\beta_I|^2\to 1$.\footnote{note that $L_N$ is invariant under the change $|\beta_I|^2\to (1-|\beta_I|^2)$} When $|\beta_I|$ approaches either of these two limiting values, the logarithmic negativity goes as 
\be
L_N = \frac{(1-2f_I)^2}{f_I(1-f_I)\ln(2)}\,  u_I^2 + \mathcal{O}(u_I^4)=\frac{2}{\ln(2)}\left(\cosh\left(\frac{\omega_I}{T}\right)-1\right) u_I^2 + \mathcal{O}(u_I^4)
\ee
where $u_I^2=|\beta_I|^2$ for the limit $|\beta_I|^2\to 0$ and $u_I^2=(1-|\beta_I|^2)$ for $|\beta_I|^2\to 1$.

\subsubsection{Optimal field for entanglement}

As in the scalar case, it is interesting to analyze the logarithmic negativity in terms of the electric field. In this context, we easily see that, at fixed $T$, there is a  value of the electric field for which the entanglement is maximal, i.e., $L_N=L_N^{(\text{max})}$.  This value is obtained for $|\beta_I|^2=1/2$. Therefore, the field that maximizes entanglement reads 
\be\label{eq:Estar}
e E_*=\frac{\pi (m^2+k_\perp^2)}{\ln 2}\, .
\ee
This critical value is independent of $T$, however, it gives smaller values of $L_N^{(\text{max})}$ as $T$ increases. For $E>E_*$ or $E<E_*$ the fermionic logarithmic negativity is always smaller than $L_N^{(\text{max})}$.

\subsubsection{Finite-resolution criterion}
We now explore benchmark values of the electric field and temperature leading to a value for the logarithmic negativity above a detectability threshold $\epsilon$, i.e. $L_N>\epsilon$. This lower bound is operationally motivated by experimental uncertainties in the construction of the density matrix. Our objective is to formulate analytical expressions that an experimenter could directly use to assess entanglement detectability, given the minimum value of logarithmic negativity that can be resolved, and determine the range of temperatures and electric fields needed for such a detection. For illustration purposes, in Appendix (\ref{apex:error in LN}), we demonstrate how a lower bound $\epsilon$ in logarithmic negativity detection can emerge given the error $\delta$ in determining the density matrix elements.\\

In what follows, we will use the following  condition for the logarithmic negativity
\be \label{eq:conditionLN}
L_N>\epsilon\, ,
\ee
with $0<\epsilon\ll1$ to define effective critical scales for $T$ and $E$ by solving $L_N=\epsilon$.
\begin{itemize}
\item {\bf Fixed $E$.} At fixed $|\beta_I|^2$ (or fixed $E$), the condition \eqref{eq:conditionLN} defines an effective critical temperature,  
 \be
T_c(\epsilon)=\frac{\omega}{\ln\left(\frac{1+2^{\epsilon/2} g(\epsilon) }{1-2^{\epsilon/2} g(\epsilon)}\right)}\,, \qquad \text{with} \qquad g(\epsilon)=\sqrt{\frac{2^{\epsilon}-1}{2^{\epsilon}-1+4(1-|\beta_I|^2)|\beta_I|^2}}\, .
\ee
Requiring $T_c(\epsilon)$ to be real and positive, namely $2^{\epsilon/2}g(\epsilon)<1$, gives $\epsilon<L_N^{(0)}$ (i.e., $\epsilon$ should be smaller than the maximum value allowed at fixed $|\beta_{I}|$, which is obtained in the zero temperature limit), with $L_N^{(0)}$ given in eq. \eqref{eq:LNT=0}. For very small $\epsilon$, we find
\be 
T_c(\epsilon)=\frac{\omega_I}{2}\frac{2^{L_N^{(0)}}-1}{\sqrt{\ln(2)\,\epsilon}} + \mathcal{O}\!\left(\epsilon\right)\, .
\ee
As expected, for $\epsilon\to0$, $T_c\to \infty$.

\item {\bf Fixed $T$}. At fixed  $T$, the condition given in eq. \eqref{eq:conditionLN}, results in the following condition for the electric field strength
\be
E_{-}(\epsilon)<E< E_{+}(\epsilon)\, ,
\ee
where 
\be 
eE_{\pm}(\epsilon)=\frac{\pi (m^2+k_\perp^2)}{-\ln|\beta_{\pm}|^2}\,, \qquad \text{with}\qquad |\beta_{\pm}(\epsilon)|^2=\frac{1}{2}\pm \frac{2^{\epsilon/2}}{2} \sqrt{1-(2^\epsilon-1)\, \text{coth}^2\left(\frac{\omega}{2T}\right)}\, . 
\ee
Imposing the square root  to be real requires $\epsilon<L_N^{(\text{max})}$, which is equivalent to demanding that the error $\epsilon$ lies below the maximum negativity at fixed $T$, which is given in eq. \eqref{eq:maximumLNf}.
For $\epsilon\to 0$, one recovers $eE_{-}\to 0$ and $e E_{+}\to\infty$, consistently with $L_N\to 0$ as $|\beta_I|^2\to 0$ or $1$.

\end{itemize}

In this sense $T_c(\epsilon)$ is the maximum temperature for which the entanglement of a given mode remains detectable (i.e.\ $L_N>\epsilon$) at fixed field strength, while $E_\pm(T,\epsilon)$ delimit the field window where entanglement remains detectable at fixed temperature. Operationally, a claim of entanglement detection requires working below $T_c$ and within the window $E_-(\epsilon)<E<E_+(\epsilon)$; outside this region the state may still be weakly entangled, but below the experimental sensitivity encoded in $\epsilon$.

\subsubsection{Axial anomaly}
As a sanity check, it is worth reproducing the axial anomaly. In the configuration $\vec E \parallel \vec B$ the Bogoliubov coefficients for Dirac fermions differ by one Landau level between the two chiral sectors, namely 
\be
|\beta^{(R)}_l|^2=\exp\,\Big[-\pi\,\frac{m^2+2leB}{eE}\Big],
\qquad
|\beta^{(L)}_{l}|^2=\exp\, \Big[-\pi\,\frac{m^2+2(l+1)eB}{eE}\Big],
\label{eq:betaRL-LL}
\ee
where we have recovered the notation introduced in Sec. \ref{sec:Schwinger}, We note that $|\beta^{(L)}_l|^2=|\beta^{(R)}_{l+1}|^2$ for all $l\ge0$.\\

For a fixed longitudinal momentum $k_z$, the mean number of created pairs in each chiral family is given in eqs. \eqref{eq:deltaNfermions}. The chiral difference summed over Landau levels then reads
\be
\sum_{l=0}^{\infty}\left(\Delta N_R-\Delta N_L\right)=\left|\beta_R(0)\right|^2 \tanh \left(\frac{\omega^{(R)}\left(0, k_z\right)}{2 T}\right) =\exp \left[-\pi \frac{m^2}{e E}\right] \tanh \left(\frac{\sqrt{m^2+k_z^2}}{2 T}\right)\, .
\ee

In the massless limit, $\Delta N_R$ and $\Delta N_L$ actually correspond to the number of right-chiral and left-chiral particles respectively so that the total axial charge by unit length reads 
\begin{align}
\Delta n_5 = \frac{2}{L}\int \frac{\text{d} k_y \text{d} k_z}{(2 \pi)^2} \sum_{l\geq 0} \left(\Delta N_R-\Delta N_L\right)
= \lim_{\Delta t\to\infty}2e B \int_0^{eE\Delta t} \frac{\text{d} k_z}{(2 \pi)^2} \tanh\left(\frac{|k_z|}{2 T}\right)
= \frac{e^2EB}{2\pi^2} \Delta t\, ,
\end{align}
where we used that for a constant electric field $E$ acting during a limited time $\Delta t$, the momentum $k_z$ is restricted to $\int \frac{d k_z}{2 \pi} \rightarrow \frac{|e E|}{2 \pi} \Delta t$, therefore 
\be
\partial_0 j_5^0=\frac{\Delta n_5}{\Delta t}= \frac{e^2}{2 \pi^2} E B \, ,
\ee
which is compatible with the axial anomaly \cite{anomaliesQFT} and agrees with the well-known independence of the anomaly at thermal equilibrium \cite{Reuter1985,Dolan:1973}. \\

 In terms of entanglement, it is interesting to note that the fermionic logarithmic negativity exhibits different magnitudes in the two chiral sectors because their effective frequencies differ by one Landau level. In particular,
\begin{equation}
(\omega^{(L)}_l)^2 = k_z^2+m^2+2(l+1)eB \;>\; (\omega^{(R)}_l)^2 = k_z^2+m^2+2leB \, .
\end{equation}
More concretely, the fermionic logarithmic negativity satisfies $L_N^{(L)}(l)=L_N^{(R)}(l+1)$ mode by mode. Therefore, when comparing the two sectors at fixed Landau level $l$, $L_N$ can be parametrically different in the $L$ and $R$ sectors for the same external parameters. Operationally, if one imposes a finite experimental threshold $L_N>\ln(1+\epsilon^2)/\ln 2$, it is possible to have entanglement above threshold in one chiral sector while it lies below threshold in the other.\footnote{See Ref. \cite{Florio:2025xup} for a detailed analysis of an entanglement asymmetry in the massless Schwinger model in 1+1.} 

\subsubsection{Mutual information}

As in the bosonic case, we can also compute the mutual information shared between particles of a given type ($L$ or $R$) and of a given mode ($I$) and its antiparticle counterpart ($\bar I$). The mutual information is defined in eq.  \eqref{eq:mutual_information} as in the scalar case, and can be computed directly from the density matrix, or from the covariance matrix using the (fermionic) Gaussian formalism. In Appendix \ref{app:gaussianforfermions} we show how this result can be computed using the Gaussian formalism for fermions, while in this subsection we use the density matrix formalism. \\

To this end, we take the defining formula of the von Neumann entropy $S_V=-\Tr[\hat \rho \log_2\hat \rho]$ and compute the appropriate traces using the explicit representation of the density matrices in a given basis (we use the {\it out}-basis $\{|00\rangle, |0 1\rangle,|1 0\rangle,|11\rangle\}$ for the two-mode system and $\{|0\rangle ,| 1\rangle\}$ for the reduced systems). The components of  $\hat\rho^{(\text{out})}_{I \bar I}$ are given in eq. \eqref{eq:out_density_matrix_fermions} while $\hat\rho^{(\text{out})}_{I }$ and  $\hat\rho^{(\text{out})}_{\bar I}$ are given in eq. \eqref{eq:reducedrhoout}. After some algebra, we get, in terms of $|\beta_I|$, $\omega_I$ and $T$ 
\be
\begin{aligned} \label{eq:Minfo-fermions}
M_{\mathrm{info}} =& \frac{2}{\ln(2)(1+e^{-\frac{\omega_I}{T}})}\left(\left(|\beta_I|^2\frac{1-e^{-2\frac{\omega_I}{T}}}{1+e^{-\frac{\omega_I}{T}}}-1\right)\ln\left( 1-|\beta_I|^2\frac{1-e^{-2\frac{\omega_I}{T}}}{1+e^{-\frac{\omega_I}{T}}}\right) \right.\\
&\left. - \left(|\beta_I|^2\frac{1-e^{-2\frac{\omega_I}{T}}}{1+e^{-\frac{\omega_I}{T}}}+e^{-\frac{\omega_I}{T}}\right)\ln\left( e^{-\frac{\omega_I}{T}}+|\beta_I|^2\frac{1-e^{-2\frac{\omega_I}{T}}}{1+e^{-\frac{\omega_I}{T}}}\right) - \frac{\omega_I}{T}e^{-\frac{\omega_I}{T}} \right)\, .
\end{aligned}
\ee
The mutual information for fermions presents some similarities and some differences with respect to the scalar case \eqref{eq:Minfo-bosons}. It is a smooth function that decreases as the thermal noise increases. However, it is not a monotonically increasing function in $|\beta_I|$, but has a maximum at $|\beta_I|^2=1/2$. As we will see in the subsequent discussion, this value of $|\beta_I|$ is associated with a value of the external electric field $E_*$ for which the mutual information (as well as the logarithmic negativity) is maximal.\\

\subsection{Connection to experiment}
\label{sec:experiment}

In vacuum QED, the relevant scale for the Schwinger effect is the Schwinger field $E_S=\frac{m_e^2 c^3}{e\hbar}\approx 1.3\times 10^{18} \mathrm{~V} / \mathrm{m}$, which corresponds to an intensity of $I_S\approx  10^{29} \mathrm{~W} / \mathrm{cm}^2$. Current state-of-the-art lasers can reach intensities up to  $I\approx 10^{23} \mathrm{~W} / \mathrm{cm}^2$ (see, for example Refs. \cite{Turcu:2016dxm,Yoon:2021ony}), i.e., $E\approx 10^{15} \mathrm{~V} / \mathrm{m}$, still several orders of magnitude below $E_S$.\footnote{Various schemes (e.g. pulse shaping or dynamical assistance) have been proposed to enhance strong-field signatures, but even optimistic projections remain below $E_S$ (see, e.g., \cite{Schutzhold:2008pz,Bulanov:2010gb,Aleksandrov:2022wcz}).} Therefore, a direct observation of vacuum pair creation remains beyond present capabilities. 
In this regime, it is also not meaningful to discuss pair entanglement: measurements capture total production and current, not correlations of specific pairs. This motivates the search for analogue realizations where the same physics can be probed under experimentally accessible conditions.\\

 In the mesoscopic Schwinger effect, an applied electric field in 
graphene induces electron-hole pair production, the solid-state analogue of vacuum pair creation. In this context, our formulas directly apply with the replacements $m c^2 \rightarrow \Delta$ (the effective gap, i.e., the minimum energy scale that appears in the channel due to its finite size) and $c \rightarrow v_F$ (the Fermi velocity), with the caveat that the experiment is effectively in $1+2$ rather than $1+3$ dimensions. This dimensionality change does not alter the mode-by-mode analysis we use, but it does affect integrated quantities such as the total current (see Ref. \cite{Kim:2000un} for explicit results in $d$ dimensions). It was recently reported universal one-dimensional 
Schwinger conductance \cite{Schmitt:2022pkd}, consistent with electron–hole pair production in graphene. For the effective gap values quoted in the experiment, namely $\Delta \lesssim 0.2 \mathrm{~eV}$, the associated Schwinger field is $E_S= \Delta^2 / e \hbar v_{\mathrm{F}}\approx 6 \times 10^{7}\mathrm{~V}/ \mathrm{m}$, which, as discussed, is experimentally accessible. The same experimental setting is also promising for probing entanglement. Focusing on the lowest-energy mode ($k_\perp=0$ and $\omega=\Delta$), the fermionic logarithmic negativity does not exhibit a sharp critical temperature, but  it is smoothly suppressed as $T/\omega$ increases. In practice, entanglement detection is limited by experimental resolution, so it is natural to adopt an operational criterion $L_N>L_N^{\text{min}}(\epsilon)$ as discussed around \eqref{eq:conditionLN} that will give an operational window for entanglement detection $E_{-}(\epsilon)<E<E_{+}(\epsilon)$ 
centered around the optimal value $E_{\star} \approx 2.7 \times 10^8 \mathrm{~V} / \mathrm{m}$. At room temperature $T=300 \mathrm{~K}$ and for $\epsilon=10^{-2}$, one finds $E_{-}\approx1.7 \times 10^7 \mathrm{~V} / \mathrm{m} $ and $E_{+}\approx 1.26 \times 10^{13} \mathrm{~V} / \mathrm{m}$ so that  $E_S$ would fit within the right window.\\

However, it is important to note that, at present, the experimental observable is the conductance, which integrates contributions from all modes. This quantity reflects the presence of pair production but is not sensitive to the quantum correlations between electron–hole partners. To access entanglement, one would need to design observables sensitive to correlations. This makes the mesoscopic platform both a confirmation of Schwinger physics and a promising starting point for future tests of its genuinely quantum character.\\

Finally, we would like to briefly mention an experimental possibility to detect the analogue Schwinger effect for bosons. In Refs. \cite{Hongo:2020xaw,Adorno:2023olb} it has been proposed that magnetic inhomogeneities in chiral magnets can trigger magnon–antimagnon pair production in close analogy to electron–positron creation in vacuum. In this mapping, the role of the electric field is played by the external magnetic profile, while magnons take the place of charged particles (i.e., the formulas above for pure electric field $E$ can be used with the replacements $eE \to \mu_{\text{eff}}\nabla B$ with $\mu_{\text{eff}}$ the effective magnetic moment of a magnon, $m_e c^2\to \Delta$ with $\Delta$ the magnon gap set by the experiment, and $c\to v_m$ with $v_m$ the spin–wave velocity). In this setting, the protocol we have outlined in subsection \ref{sec:initialsqueezed} to maximize entanglement through squeezed initial states could also be implemented. In fact, proposals to engineer squeezed magnon states already exist in related cavity platforms~\cite{Li:2019cxy}, suggesting that this route might provide a realistic option to test  quantum correlations in a bosonic analog of the Schwinger mechanism.

\section{Conclusions}
\label{sec:conclusions}

In this article, we have analyzed pair creation in strong-field QED from a quantum-information perspective, focusing on entanglement for scalar and spinor fields in constant backgrounds. Working mode by mode, and using Gaussian tools when required, we derived closed-form expressions for particle production, logarithmic negativity and mutual information starting from thermal initial states.\\

For bosons, thermal fluctuations enhance particle production through the Bose factor $\coth (\omega/2T)$ (see eq. \eqref{eq:differenceN-bosons}). However, the quantum character of the process --captured by the logarithmic negativity-- survives only below a (mode-dependent) critical temperature $T_c$ as we show in eq. \eqref{eq:Tc-bosons}. Analogously, at a fixed temperature, we found that the bosonic pairs are entangled only if $E>E_{\text{entang}}$ with the critical field given in eq.\eqref{eq:Ecrit-entanglement-bosons}. We also discussed how to enhance entanglement by means of an initial squeezed state.\\

 For fermions, Pauli blocking competes with pair creation, producing the characteristic $\tanh(\omega/2T)$ in the particle number (see eq. \eqref{eq:deltaNfermions}). Entanglement exhibits a qualitatively different pattern with respect to the scalar case: 
 the logarithmic negativity is a smooth function of $T$ and $|\beta_I|^2$ (which is constrained between 0 and 1). It is monotonically suppressed as $T$ increases, and, at fixed $T$, it reaches its maximum at $\left|\beta_I\right|^2=1/2$, which defines a temperature-independent optimal field $E=E_*$ (see eq.  \eqref{eq:Estar}). For experimental purposes, one may nonetheless introduce an operational thereshold $L_N>\epsilon$, which induces an effective entanglement window $E_{-}(\epsilon)<E< E_{+}(\epsilon)$ and a corresponding temperature scale $T_c(\epsilon)$ (see eq. \eqref{eq:conditionLN} and the subsequent discussion). As a consistency check, we reproduced the axial anomaly and commented on the entanglement differences between the two sectors ($L$ and $R$).  \\

Finally, we discussed the implications for near and mid-term experiments. In vacuum QED, the Schwinger scale $E_S=m^2_ec^3/e\hbar$ is not yet accessible, but analogue experiments operate in regimes where our mode-by-mode criteria and thresholds are, in principle, applicable. Beyond constant fields, our formulas could be applied directly to pulsed profiles once the corresponding Bogoliubov coefficients are given.

\begin{acknowledgments}
The authors thank Björn Garbrecht, Pablo Chisvert, and Ivan Agullo  for their insightful comments.  The work of SP was funded by the Deutsche Forschungsgemeinschaft (DFG, German Research Foundation) under Germany’s Excellence Strategy – EXC 2094 – 390783311. DK acknowledges financial support by DIM ORIGINES program from {\textit{R\'egion \^Ile de France}}. DK also received support through the Atracci\'{o}n de Talento Cesar Nombela grant No 2023-T1/TEC-29023, funded by Comunidad de Madrid (Spain); as well as financial support via the Spanish Grant PID2023-149560NB-C21, funded by MCIU/AEI/\allowbreak 10.13039/\allowbreak 501100011033/\allowbreak FEDER, UE. D.~K. also received support through Plan GenT2 (DEGENT) Grant No.~CIDEXG/2026/54, funded by Generalitat Valenciana (Spain), and the WOST (WithOut SpaceTime) project~\cite{WOST}, led by the Center for Spacetime and the Quantum (CSTQ), and supported by Grant ID~63683 from the John Templeton Foundation (JTF). The opinions expressed in this publication are those of the authors and do not necessarily reflect the views of the Center for Spacetime and the Quantum.
\end{acknowledgments}

\appendix

\section{Details for fermions} \label{ap:fermionsB0}

In this appendix, we give some details for fermions that are complementary to the analysis in Section \ref{sec:Schwinger}.

\subsection{Spinor solutions for $B=0$}
Here we study in detail the spinor solutions for the $\vec B=0$ case. The mode expansion given in \eqref{eq:mode-expansion-fermions-BE} can be still used with the replacement
 \be
 \int \text{d}^2\,k \,e^{i\vec{k}\cdot\vec{y}}\sum_{l\in\mathbb{N}}e^{-\frac{\eta^2}{2}}H_l(\eta) \to \int \text{d}^3k \, e^{i \vec k \vec x}\, , 
 \ee
and  the variables
\be
u=\frac{k_z +eEt}{\sqrt{eE}}\, ,\qquad  \lambda_k= \frac{k_x^2 + k_y^2 +m^2}{eE}\, .
\ee
Following the same strategy as in the $\vec B \parallel \vec E$ case, it is possible to find two basis of solutions that are well behaved in the asymptotic past and future respectively,
\be
\mathcal{B}^{\mathrm{in}} = \Big\{ U_{\mathrm{in}}^L, U_{\mathrm{in}}^R, V_{\mathrm{in}}^L, V_{\mathrm{in}}^R\Big\} ~~~\mathrm{and}~~~ \mathcal{B}^{\mathrm{out}} = \Big\{ U_{\mathrm{out}}^L, U_{\mathrm{out}}^R, V_{\mathrm{out}}^L, V_{\mathrm{out}}^R\Big\}\, \, ,
\ee
where 
\bea
U_{\mathrm{in/out}}^L(\vec{k},u) &=&  \mathscr{N}_{\lambda_k}\,L_{1/2}(\vec k,u) \, ,\qquad \quad \,\,\, \quad U_{\mathrm{in/out}}^R(\vec{k},u) =  \mathscr{N}_{\lambda_k} \,R_{1/2}(\vec{k},u)\, , \\
V_{\mathrm{in/out}}^L(\vec{k},u) &=& \sqrt{2}\mathscr{N}_{\lambda_k}\, L_{-2/-1}(\vec{k},u) \,, \qquad V_{\mathrm{in/out}}^R(\vec{k},u) = \sqrt{2}\mathscr{N}_{\lambda_k}\,R_{-2/-1}(\vec{k},u)\, ,
\eea
and with the normalization constant being
\be
\mathscr{N}_{\lambda_k}=\frac{1}{(2\pi)^{3/2}} \frac{e^{-\frac{\lambda_k\pi}{8}}}{\sqrt{4eE}}\, ,
\ee
and where 
\bea
L_{i}(\vec{k},u) &=& (A_{L}(\vec{k}) + B_{L}(i\partial_u-u))S_{i}(\lambda_k,u)\, ,\\
R_{i}(\vec{k},u) &=& (A_{R}(\vec{k}) + B_{R}(i\partial_u-u))S_{i}(\lambda_k,u)\, ,
\eea
with $i=\{\pm1,\pm2\}$ and 
\be
A_{L}(\vec{k}) = \begin{pmatrix}m\\-k_x-ik_y\\m\\k_x+ik_y\end{pmatrix} ~;~~~B_{L} = \begin{pmatrix}\sqrt{eE}\\0\\-\sqrt{eE}\\0\end{pmatrix} ~;~~~ A_{R}(\vec{k}) = \begin{pmatrix}k_x-ik_y\\m\\k_x-ik_y\\-m\end{pmatrix} ~;~~~B_{R} = \begin{pmatrix}0\\\sqrt{eE}\\0\\\sqrt{eE}\end{pmatrix}\, .
\ee
To find the relationship between the {\it in} and {\it out} bases we follow the strategy given in subsection \ref{subsec:Bogoliubov}, and find exactly eqs. \eqref{Bogoliubov_coeffs_schwinger_spinor_u} and \eqref{Bogoliubov_coeffs_schwinger_spinor_v} with the replacements $\lambda_l\to \lambda_k$ and $\lambda_{l+1}\to \lambda_k$.

\subsection{Vacuum persistence amplitude from the Bogoliubov coefficients}
As an illustration, we show here how the Schwinger formula can be obtained from the Bogoliubov coefficients for spinors in the case $\vec B\parallel \vec E$.
The Schwinger effect manifests itself in the probability of observing the vacuum state at late times when the system has been prepared in the vacuum state at early times. If the vacuum persistence probability  is different (and less than) one, it means that particles have been created out of the vacuum.\\

Since we work with fermions, the average number of particles equals the probability of producing a pair $\langle N_I\rangle=|\beta_I|^2=P(1)$. Therefore we can compute the vacuum-to-vacuum probability as $P(0)=1-P(1)$ for each mode (for more details see, for example, Ref. \cite{parker-toms}, chapter 2). Therefore, the vacuum persistence amplitude (taking into account all modes) reads
\begin{align*} 
|{}_{\mathrm{in}}\langle 0|0\rangle_{\mathrm{out}}|^2 &= \prod_{\vec{k}}\prod_l (1-|\beta^{(R)}_{\lambda_l}|^2)(1-|\beta^{(L)}_{\lambda_l}|^2)\\
&= \exp\left( \frac{L^2}{(2\pi)^2}\int \text{d}^2k \sum_{l\in\mathbb{N}} \left( \ln(1-e^{-\lambda_{l}\pi})+\ln(1-e^{-\lambda_{l+1}\pi}) \right)\right) \\
&= \exp\left( \frac{L^2}{(2\pi)^2}\int \text{d}^2k \sum_{l\in\mathbb{N}} \sum_{n=1}^{+\infty} \frac{-1}{n}(e^{-n\lambda_l\pi}+e^{-n\lambda_{l+1}\pi})\right) \\
&= \exp\left( -\frac{L^2}{(2\pi)^2}e^2|EB|\int \text{d}x \text{d}t \sum_{n=1}^{+\infty} \frac{1}{n}\exp\left(-n\pi\frac{m^2}{eE}\right)\coth\left(n\pi\frac{B}{E}\right)\right) \\
&= \exp\left( -\frac{V\Delta t}{(2\pi)^2}e^2|EB| \sum_{n=1}^{+\infty} \frac{1}{n}\exp\left(-n\pi\frac{m^2}{eE}\right)\coth\left(n\pi\frac{B}{E}\right)\right)
\end{align*}
where we used the changes of variables $\text{d}k_y = -eB\text{d}x$ and $\text{d}k_z = -eE\text{d}t$ and thus $\int \text{d}k_y = |eB| L$ and $\int \text{d}k_z = |eE|\Delta t$ where $L = V^{1/3}$. This  result is in agreement with the standard formula \cite{Popov:1971iga, Dunne:2004nc}.

\section{Gaussian formalism for fermions/details for fermionic systems}
\label{app:gaussianforfermions}

Although the fermionic Gaussian formalism is not needed for the observables discussed in this work, since all relevant quantities can be directly computed  from the finite density matrix (see Section \ref{sec:twolevel}), it provides a compact and parallel way to treat bosons and fermions in the same language. We will give here an operational approach. For a formal discussion, see, for example, Ref. \cite{Hackl:2020ken}. As in the scalar case, for a pair of fermionic modes, their respective ladder operators $\{\hat b_1,\hat b^\dagger_1, \hat b_2, \hat b^\dagger_2\}$ can be arranged in a 4-vector $\bm \psi_J$. After interaction, in the Heisenberg picture, the $^{(\text{in})}$ and $^{(\text{out})}$ operators are related by a Bogoliubov transformation:
\be \label{eq:bogo_fermions_gaussian}
\bm{\psi}_J=\begin{pmatrix} \hat b_{1,J} \\ \hat b^\dagger_{1,J} \\ \hat{b}_{2,J} \\ \hat{b}^\dagger_{2,J} \end{pmatrix}\, ,\qquad  \bm{\psi}_J^{(\text{out})}= \bm B_J \bm{\psi}_J^{(\text{in})} \qquad \bm{B}_J=
\begin{pmatrix}
\alpha_J&0& 0 &\beta_J\\
0&\alpha^*_J&\beta^*_J&0\\
0&-\beta_J&\alpha_J&0\\
-\beta^*_J&0&0& \alpha^*_J
\end{pmatrix}\, .
\ee
It is often assumed that $\alpha$ is real, so that the Bogoliubov coefficients are parametrized as 
\be \alpha=\cos s\,, \qquad   \beta=e^{i\varphi}\sin s\, .
\ee
where $s\in (0,\pi/2)$. As in the scalar case, it is convenient to work with the vector of operators $\hat {\bm R}=(\hat x_1, \hat p_1,\hat x_2, \hat p_2)^\top$, where $\hat x$ and $\hat p$ are defined as in \eqref{eq:xpa_tranformation}. Contrary to the scalar case, $\bm R$ now satisfies the anticommutation relation 
\begin{equation}
\{\hat {\bm{R}}, \hat{\bm {R}}^{\top}\} = {\bm I}_{4}
\end{equation}
where ${\bm I}_{4}$ is the $4\times 4$ identity matrix. The first moment ${\bm \mu}$ and the covariance matrix ${\bm \sigma}$ are accordingly defined by
\begin{equation}
\begin{dcases}
{\bm\mu} = \langle {\bm{ \hat{R}}}\rangle \\
{\bm \sigma} = \langle [{\bm{ \hat{R}}}-\bm{\mu}, { \hat{\bm{R}}^{\top}}-\bm{\mu^{\top}}]\rangle\, .
\end{dcases}
\end{equation}
We note that for physical states $\bm\mu=0$ as discussed, for example, in Ref. \cite{Shapourian:2018ozl}.\\

 For our purposes, we focus on initial thermal states. It is not hard to show that the first and second moments for each two-mode sector $(I\bar I)$ read
 \be
 \label{eq:instatefermions}
\bm{\mu}_{I\bar I}^{(\mathrm{in};\,i)}=\bm{0}\, , \qquad \text{and} \qquad \bm{\sigma}_{I\bar I}^{(\mathrm{in};\, i)}=i(1-2f^{(i)}_I)\bm{\Omega}_4
\, ,\qquad \text{with}\qquad f^{(i)}_I=\frac{1}{e^{\omega^{(i)}_I/T}+1}\, .
\ee
where $\bm\Omega_4$ is the fermionic symplectic form and with $i=(R,L)$. From now on, we will omit the superindex $^{(i)}$ except when necessary. We can obtain the first two moments of the {\it out} state from the Bogoliubov transformation
\be
\bm\mu^{(\mathrm{out})}_{I\bar I}=\bm O\,\bm\mu^{(\mathrm{in})}_{I\bar I}\,, \qquad \bm\sigma^{(\mathrm{out})}_{I\bar I}=\bm O\,\bm\sigma^{(\mathrm{in})}_{I\bar I}\bm O^{\, \top},
\ee
where\footnote{$\bm O$ is derived from \eqref{eq:bogo_fermions_gaussian} by simply applying a standard change of basis $\{\hat b,\hat b^\dagger\}\to \{\hat x,\hat p\}$.} 
\begin{equation}
\bm{O}=
\begin{pmatrix}
\cos s&0&\cos \varphi \sin s&\sin \varphi \sin s\\
0&\cos s&\sin \varphi \sin s&-\cos \varphi \sin s\\
-\cos \varphi \sin s&-\sin \varphi \sin s&\cos s&0\\
-\sin \varphi \sin s&\cos \varphi \sin s&0& \cos s
\end{pmatrix}, \label{eq:Osq}
\end{equation}
Therefore the relationship between $s$ and $\varphi$ and our Bogoliubov coefficients given in eqs. \eqref{Bogoliubov_coeffs_schwinger_spinor_u} (for type $L$) and \eqref{Bogoliubov_coeffs_schwinger_spinor_v} (for type $R$) is\footnote{In our computation for the Schwinger effect $\beta$ is real and $\alpha$ is complex. While in the standard derivation it is the other way around. This is not a problem, since the coefficients are related by a simple rotation that does not change the final result.}
\be
\cos s =|\alpha_I|\, , \qquad \sin s = |\beta_I| \, , \qquad \varphi=\arg(\beta_I \alpha^*_I)\, .
\ee
 It is not difficult to see that $\bm\sigma^{(\mathrm{out})}_{I}=\bm\sigma^{(\mathrm{out})}_{\bar I}=i(1-2f_I)\cos(2s)\bm\Omega_2$. The average number of particles (and of antiparticles) of a given type ($R$ or $L$) and in a given mode $I$ ($\bar I$) can be obtained using $\langle\hat N_I^{(\mathrm{out})}\rangle =[1+i(\bm{\sigma}_I^{(\mathrm{out})})_{12}]/2$. It gives
\begin{equation}
\langle \hat{N}^{(\mathrm{out})}_I\rangle = |\beta_I|^2 + \frac{1-2|\beta_I|^2}{1+e^{\omega_I/T}} = f_I + |\beta_I|^2\left(1-2f_I\right)=\frac{1}{2}\left(1-(1-2f_I)\cos 2s \right)\, .
\label{eq:ap:averagefermions}
\end{equation}

We can also compute the mutual information using the Gaussian formalism. To this end, we first write the von Neumann entropy in terms of the positive eigenvalues of the covariance matrix $\bm \sigma$, as we did for the scalar case. We use $S_V=\sum_J h(\nu_J)/\ln(2)$ for the von Neumann entropy, with $h$ defined below and eq. \eqref{eq:mutual_information} for the mutual information, with the only difference  that now $\nu_J$ represent an eigenvalue of the covariance matrix $\bm\sigma$ itself, and not of the symplectic one ($i\bm{\Omega \sigma}$). Let us denote $\{\pm \nu_{I}\}$ the two eigenvalues of $\bm \sigma_I$  and $\{\pm \nu_{I \bar I}\}$ the four eigenvalues of $\bm \sigma_{I\bar I}$, each of them with multiplicity two. For the {\it out} state, we obtain that the relevant (positive) eigenvalues are 
\be
\nu_{I\bar I}= (1-2f_I)\,\qquad  \nu_I=\nu_{\bar I}=(1-2f_I)\cos(2s)
\ee
Hence, the mutual information reads 
\be \label{eq:mutualinfo_fermions_short}
M_{\text{info}}=\frac{2}{\ln 2}\left(h((1-2f_I)\cos(2s))-h(1-2f_I)\right)\, .
\ee
with $h(x)=-\frac{1+x}{2} \ln \frac{1+x}{2}-\frac{1-x}{2} \ln \frac{1-x}{2}$, with $0\ln 0=0$. The same result is obtained directly from the density matrix, as we show in the main text \eqref{eq:Minfo-fermions}.

\section{Operational lower bound in Logarithmic Negativity} \label{apex:error in LN}

In this appendix, we derive the error in determining the logarithmic negativity given the experimental error $\delta$ in the construction of the density matrix. \\

Consider a $d\times d$ density matrix $\hat{\rho}$. Let $\hat{\rho}'$ be the reconstructed density matrix from an experiment that is supposed to match $\hat{\rho}$. Let $\delta_{ij}$ be the error bars of the (theoretically) expected density matrix elements, i.e.
\begin{equation}
|\hat{\rho}_{ij}-\hat{\rho}_{ij}'|\leq \delta_{ij}, \quad \text{for}\, (i,j)\in S. \label{eq:rho_error}
\end{equation}
where $S=\{(i,j)\in \{1,...,d\}^2:\hat{\rho}_{ij}\neq 0\}$, and let $s\leq d^2$ be the number of the non-zero elements of $\hat{\rho}$ (and $\hat{\rho}'$). For simplicity, we assume that $\hat{\rho}'_{ij}=0$ whenever $(i,j)\notin S$.\\

The logarithmic negativity $L_N$, measuring the entanglement across a partition in the state $\hat{\rho}$, is defined as
\begin{equation}
L_N(\rho)=\log_2\!\left\|\hat{\rho}^{\text{PT}}\right\|_1\, .
\label{eq:LN_defin}
\end{equation}
 Here, $\hat{\rho}^{\text{PT}}$  represents the partially transposed density matrix for bosons, and the partially time reversed density matrix for fermions.  $||\hat{X}||_1:=\Tr \sqrt{\hat{X}^\dagger\hat{X}}$ is the trace norm of the operator $\hat{X}$. \\

We are interested in the error in logarithmic negativity, i.e. we want to find an upper bound in 
\begin{equation}
\left|L_N(\hat{\rho})-L_N(\hat{\rho}')\right|=\left|\log_2||\hat{\rho}^{\text{PT}}||_1-\log_2||\hat{\rho}^{'\text{PT}}||_1\right|=\frac{1}{\ln 2}\left|\ln||\hat{\rho}^{\text{PT}}||_1-\ln||\hat{\rho}^{'\text{PT}}||_1\right|. \label{eq:LN_diff}
\end{equation}
Notice that from (\ref{eq:rho_error}) and assuming, for simplicity, a uniform error across the non-zero entries, $\delta_{ij}=\delta$, we can write
\begin{equation}
|\hat{\rho}^{\text{PT}}_{ij}-\hat{\rho}_{ij}^{'\text{PT}}|\leq \delta\, . \label{eq:rhoPT_error}
\end{equation}
 
To start, it is easier to compute the Frobenius norm of the partially transposed density matrix difference and then relate this to the corresponding trace norms. The Frobenius norm of an operator $\hat{X}$ with matrix entries $X_{ij}$ is
\begin{equation}
||\hat{X}||_2:=\sqrt{\sum_{i,j}|X_{ij}|^2}.
\end{equation}
Setting $\hat{X}=\hat{\rho}^{\text{PT}}-\hat{\rho}^{'\text{PT}}$
\begin{equation}
||\hat{\rho}^\text{PT}-\hat{\rho}^{'\text{PT}}||_2:=\sqrt{\sum_{i,j}|\hat{\rho}_{ij}^{\text{PT}}-\hat{\rho}^{'\text{PT}}_{ij}|^2} \leq \sqrt{s} \delta, \label{eq:Frobenius_norm_rho_diff}
\end{equation}
where we made use of the fact that $\hat{\rho}$ and $\hat{\rho}^{'\text{PT}}$ have $s$ non-zero entries. \\

Now, we want to connect the Frobenius to the trace norm since it is the latter that enters in the definition of logarithmic negativity (\ref{eq:LN_defin}). To this end, we use the inequality (3) of reference \cite{Coles_2019}
\begin{equation}
||\hat{X}||_1\leq \sqrt{ \text{rank}(\hat{X})} ||\hat{X}||_2.
\end{equation}
For $\hat{X}=\hat{\rho}^{\text{PT}}-\hat{\rho}^{'\text{PT}}$, with $\text{rank}(\hat{\rho}^{\text{PT}}-\hat{\rho}^{'\text{PT}})\leq d$, we get
\begin{equation}
||\hat{\rho}^{\text{PT}}-\hat{\rho}^{'\text{PT}}||_1\leq \sqrt{d}||\hat{\rho}^{\text{PT}}-\hat{\rho}^{'\text{PT}}||_2. \label{eq:trace_to_Frobenius_norms}
\end{equation}
Thus, from (\ref{eq:Frobenius_norm_rho_diff}) and (\ref{eq:trace_to_Frobenius_norms}), one finds
\begin{equation}
||\hat{\rho}^{\text{PT}}-\hat{\rho}^{'\text{PT}}||_1 \leq \sqrt{sd}\,\delta \label{eq:trace_norm_upper_bound}
\end{equation}
Furthermore, the several norms obey the triangle inequality. For instance, for the trace norm of two operators $\hat{X}, \hat{Y}$, we have
\begin{equation}
||\hat{X}+\hat{Y}||_1\leq ||\hat{X}||_1+||\hat{Y}||_1
\end{equation}
Setting $\hat{X}=\hat{\rho}^{\text{PT}}$ and $\hat{Y}=\hat{\rho}^{'\text{PT}}-\hat{\rho}^{\text{PT}}$, we find
\begin{equation}
||\hat{\rho}^{'\text{PT}}||_1-||\hat{\rho}^{\text{PT}}||_1 \leq ||\hat{\rho}^{\text{PT}}-\hat{\rho}^{'\text{PT}}||_1
\end{equation}
Similarly, setting $\hat{X}=\hat{\rho}^{'\text{PT}}$ and $\hat{Y}=\hat{\rho}^{\text{PT}}-\hat{\rho}^{'\text{PT}}$, we find
\begin{equation}
||\hat{\rho}^{\text{PT}}||_1-||\hat{\rho}^{'\text{PT}}||_1 \leq ||\hat{\rho}^{\text{PT}}-\hat{\rho}^{'\text{PT}}||_1
\end{equation}
Combining the last two inequalities and applying (\ref{eq:trace_norm_upper_bound}), one finds
\begin{equation}
\left|||\hat{\rho}^{\text{PT}}||_1-||\hat{\rho}^{'\text{PT}}||_1\right| \leq ||\hat{\rho}^{\text{PT}}-\hat{\rho}^{'\text{PT}}||_1\leq \sqrt{sd}\delta. \label{eq:trace_norm_diff_upper_bound}
\end{equation}
Furthermore, one can show, e.g. using the mean value theorem, that for two real numbers $a$, $b$, satisfying $1\leq a \leq b$,
\begin{equation}
\ln b-\ln a\leq b-a, \label{eq:MVT}
\end{equation}
Thus, setting $a=||\hat{\rho}^{\text{PT}}||_1$ and  $b=||\hat{\rho}^{'\text{PT}}||_1$, expression (\ref{eq:MVT}) applied to (\ref{eq:LN_diff}) results in
\begin{equation}
\left|L_N(\hat{\rho})-L_N(\hat{\rho}')\right|\leq\frac{1}{\ln 2}\left|||\hat{\rho}^{\text{PT}}||_1-||\hat{\rho}^{'\text{PT}}||_1\right|. \label{eq:LN_diff_aux}
\end{equation}
Finally, combining (\ref{eq:trace_norm_diff_upper_bound}) with (\ref{eq:LN_diff_aux}), we obtain
\begin{equation}
\left|L_N(\hat{\rho})-L_N(\hat{\rho}')\right|\leq \frac{\sqrt{sd}\delta}{\ln 2}. \label{eq:LN_diff_final}
\end{equation}
We define the threshold parameter
\begin{equation}
\epsilon(\delta):=\frac{\sqrt{sd}}{\ln2}\delta. \label{eq:epsilon}
\end{equation}
Treating $L_N(\hat{\rho})$ as the theoretically expected value and $L_N(\hat{\rho}')$ as the experimentally reconstructed one, we require
\begin{equation}
L_N(\hat{\rho})>\epsilon(\delta)
\end{equation}
to guarantee that $L_N(\hat{\rho}')>0$. Hence, $\epsilon(\delta)$ provides the detection threshold for the theoretically expected logarithmic negativity.\\

For the \textit{out} density matrix in the fermionic Schwinger effect (\ref{eq:out_density_matrix_fermions}), there are only $6$ non-zero elements of $\hat{\rho}$. Plugging $d=4$, $s=6$, we obtain
\begin{equation}
L_N(\hat{\rho})> \frac{2\sqrt{6}}{\ln 2}\delta\approx 7.07 \delta.
\end{equation}

\bibliography{refs_final}

\end{document}